\documentclass[a4paper, 11pt]{article}
\pdfoutput=1
\usepackage{jheppub}

\title{The 10d Uplift of the GPPZ Solution}
\author[a]{Michela Petrini,}
\author[b]{Henning Samtleben,}
\author[c]{Stanislav Schmidt,}
\author[c]{Kostas Skenderis}
\affiliation[a]{Sorbonne Universit\'e, CNRS, Laboratoire de Physique Th\'eorique et Hautes \'Energies, F-75005 Paris, France}
\affiliation[b]{Univ Lyon, Ens de Lyon, Univ Claude Bernard, CNRS,
Laboratoire de Physique,\\ F-69342 Lyon, France.}
\affiliation[c]{
	STAG Research Centre and Mathematical Sciences, University of Southampton,\\
	Southampton, SO17 1BJ, UK}
\emailAdd{Petrini@lpthe.jussieu.fr}
\emailAdd{Henning.Samtleben@ens-lyon.fr}
\emailAdd{S.Schmidt@soton.ac.uk}
\emailAdd{K.Skenderis@soton.ac.uk}

\abstract{We present the uplift of the GPPZ solution of the five-dimensional
maximal supergravity to ten dimensions. The five dimensional solution involves
two real scalar fields, with one of them encoding holographically the (norm of
the complex) supersymmetric ${\mathcal N}=1$ mass deformation and the other the
real part of the gaugino condensate. We embed this solution in a consistent
truncation of $D=5$ maximal supergravity which involves two complex scalars dual
to the complex mass deformations and the complex gaugino condensate, and a
$\textup{U}(1)$ gauge field dual to the $\textup{U}(1)_R$ current, and uplift it
to ten dimensions. The ten dimensional solution is completely explicit, with all
fields given in terms of elementary functions. The metric and the axion-dilaton
agree with those of a partial uplift of the GPPZ flow by Pilch and Warner. We
analyze the asymptotics and the singularity structure of the ten dimensional
solution. The uplifted solution is singular, but the singularity is milder than
that of the five dimensional solution, and there is conformal frame in which the
metric is only singular at one point of $S^5$. We compare the asymptotics of the
$10d$ solution with that of the Polchinski-Strassler and Freedman-Minahan
solutions, and find agreement with Freedman-Minahan and disagreement with
Polchinski-Strassler. In particular, we infer that while the
Polchinski-Strassler $10d$ fields satisfy the correct boundary conditions, they
do not solve the field equations near the boundary. }

\usepackage[utf8x]{inputenc} % To be able to type letters with accents directly, e.g. Poincaré
\usepackage[open,numbered]{bookmark}
\usepackage{tensor}
\allowdisplaybreaks{}

\begin{document}
\maketitle

%%%%%%%%%%%%%%%%%%%%%%%%%%%%%%%%%%%%%%%%%%%%%%%%%%%%
%
% Introduction
%
%%%%%%%%%%%%%%%%%%%%%%%%%%%%%%%%%%%%%%%%%%%%%%%%%%%%
\section{Introduction and Summary of Results}%
\label{sec:introduction}

Since the early days of the AdS/CFT correspondence many efforts have been
devoted to constructing gravity duals of $\mathcal{N}=1$ gauge theories in four
dimensions. While we have now a very good understanding of the duality for
superconformal $\mathcal{N}=1$ theories, the same is not true for non-conformal
ones. The two best known $\mathcal{N}=1$ solutions, the
Maldacena-Nunez~\cite{Maldacena:2000yy} and the
Klebanov-Strassler~\cite{Klebanov:2000hb} ones, are dual to gauge theories with
unconventional UV completions, namely higher-dimensional theories or theories
with an infinite number of degrees of freedom. A natural theory to be studied
holographically is $\mathcal{N}=1^*$. This is obtained as a deformation of
$\mathcal{N}=4$ Super Yang Mills and, according to the AdS/CFT dictionary, its
supergravity dual should correspond to a deformation of $AdS_5 \times S^5$.
However the issue of finding such solution is still not settled.

The $\mathcal{N}=1^*$ theory is obtained by adding a mass term for the three
chiral superfields of $\mathcal{N}=4$ Super Yang Mills. This amounts to adding
to the $\mathcal{N}=4$ superpotential the term \begin{equation}
\label{eq:massterm} \delta \mathcal{W} = m_{ij} {\rm tr} (\Phi_i \Phi_j) \;,
\end{equation} which reduces the supersymmetry to $\mathcal{N}=1$ and breaks
explicitly the R-symmetry. At energies lower than the mass scale the chiral
matter multiplets decouple, the $\textup{U}(1)$ R-symmetry is recovered and the
theory flows to pure $\mathcal{N} = 1$ Super Yang Mills. $ \mathcal{N} = 1$
Super Yang Mills confines and has a mass gap, with $N$ different vacua
associated with the gaugino condensates.

Far from the decoupling limit, the $\mathcal{N}=1^*$ theory has a rich structure
of vacua as described in~\cite{Vafa:1994tf, Donagi:1995cf, Polchinski:2000uf}.
Some vacua are characterised by a mass gap, which can be due to a Higgs
mechanism or to confinement, while some others contain massless photons and
hence have no mass gap. At the classical level the vacua are parameterised by
the $N$ dimensional representations of $\textup{SU}(2)$, as one can see from the
F-term equation\footnote{ We consider the case where the mass term is diagonal.
By rescaling the superfields $\Phi_i$ one can always write the mass term as
$m_{ij} = m \delta_{ij}$.}
\begin{equation}
\label{Ftermd}
[\Phi_i , \Phi_j ] = - \frac{m}{\sqrt{2}} \epsilon_{ijk} \Phi_k \, .
\end{equation}
The $N$-dimensional irreducible representation of $\textup{SU}(2)$ corresponds
to the Higgs vacuum. The gauge group is completely broken and there is a mass
gap already at the classical level. As the theory is weakly coupled at all
energy scales, the semiclassical analysis holds also at the quantum level and
there is exactly one vacuum. The opposite case, corresponding to $N$ copies of
the trivial representation, is the $\textup{SU}(N)$ confining vacuum. All the
vevs of the scalars are zero and the gauge group is unbroken. At the quantum
level the theory confines in the IR and the classical vacuum splits into $N$
vacua parameterised by the gaugino condensate $\langle \lambda \lambda \rangle$.
Since the R-symmetry is explicitly broken already in the UV, the different vacua
are not related by a discrete R-symmetry as in pure $\mathcal{N} = 1$ Super Yang
Mills, and are not isomorphic. In each vacuum the superpotential takes different
values and the domain walls connecting the vacua have different tension. This is
the vacuum that survives in the decoupling limit.

Other confining vacua appear when the fields $\Phi_i$ consist of blocks that are
all in the same representation of dimension $p$. The residual gauge group is
$\textup{SU}(p)$. At the quantum level the theory confines and there are $p$
vacua parameterised by the gaugino condensates. If, on the contrary, the
$\Phi_i$ split in blocks of the different dimensions, the vacua will generically
have $\textup{U}(1)$ factors and, hence, massless photons. These vacua have no
mass gap and are called Coulomb vacua.

A first attempt to determine the gravity dual of the $\mathcal{N}=1^*$ theory is
the so called GPPZ solution~\cite{Girardello:1999bd} in $\mathcal{N}=8$
$\textup{SO}(6)$ gauged supergravity in five-dimensions~\cite{Pernici:1985ju,
Gunaydin:1984qu}. This is a consistent truncations of type IIB supergravity on
$AdS_5 \times S^5$~\cite{Baguet:2015sma} that keeps only the lightest modes of
the Kaluza-Klein reduction on $S^5$~\cite{Kim:1985ez}. On the gauge theory side
these modes correspond to the relevant operators in $\mathcal{N}=4$ Super Yang
Mills and thus contain the mass deformations. By imposing that the masses of
three chiral superfields are the same it is possible to consistently truncate
the five-dimensional supergravity to only two scalars, $m$ and $\sigma$, and to
find an analytic solution of the five-dimensional equations of motion. The
asymptotic behaviour of the two scalars confirm that $m$ and $\sigma$ can be
identified with a mass term for the matter superfields and the gaugino
condensate, respectively.

The solution has a naked singularity. Nevertheless it is possible to perform
several computations on the solution and obtain sensible results that seem to
confirm its interpretation as the dual of
$\mathcal{N}=1^*$.\footnote{In~\cite{Aharony:2000nt} it was argued that there
cannot be a supergravity description of the confining vacua of ${\cal N}=1^*$
theory. This claim was based on the structure of the vacuum expectation values
at strong coupling, which was obtained using an alleged modular symmetry of the
chiral sector of the theory in each vacuum. The existence of this symmetry
relied on a contribution of the Konishi anomaly to the ${\bf 10}$ and ${\bf
\overline{10}}$, which is now known not to be correct. In more detail, equation
(2.4) of~\cite{Aharony:2000nt} contains a quantum contribution to the ${\bf
10}$. The operator in the ${\bf 10}$ however is a supersymmetric operator and as
such it cannot mix with the non-supersymmetric Konishi operator (and this has
been checked explicitly to 2-loop order in~\cite{Eden:2005ve}). This invalidates
equation (4.9) in~\cite{Aharony:2000nt} and consequently the derivation of the
vevs at strong coupling. We thank Ofer Aharony for a discussion about this
point.} It can be easily checked that the theory admits a mass gap and a
discrete spectrum of glueballs~\cite{Petrini:1999qa}. Moreover, the holographic
computation of 2-point functions along the flow gave results consistent with
interpretation of the solution as a deformation of ${\cal N}=4$ SYM, including
subtle issues regarding the analytic structure of the
correlators~\cite{Bianchi:2001de, Bianchi:2001kw}. The GPPZ solution admits two
truncations, with $\sigma =0$ and with $m=0$, respectively. For the $m=0$
solution the spectrum of light scalars has been computed
in~\cite{Elander:2010wd} and appears to be insensitive to the IR singularity. On
the other hand,~\cite{DeWolfe:2000xi} found a massless state in the spectrum of
theory, which may point towards the interpretation of the solution as being dual
to a Coulomb vacuum. Even if the solution has many qualitative features
consistent with being dual of one of the confining vacua of $\mathcal{N}=1^*$,
due to the naked singularity, its physical interpretation is still not
completely clear. Sometimes singularities are resolved when a solution is
uplifted in higher dimensions~\cite{Gibbons:1994vm}. For example, the solutions
describing the Coulomb Branch of ${\cal N}=4$ SYM have a naked singularity from
the $5d$ perspective but they are non-singular from the $10d$
perspective~\cite{Freedman:1999gk}. Therefore, the first step towards assessing
the validity of the GPPZ solution is to uplift it to 10 dimensions and this is
the topic of this paper.

The general uplift of $\mathcal{N}=8$ $\textup{SO}(6)$ gauge supergravity to
type IIB was constructed in~\cite{Baguet:2015sma}. Here our starting point is a
4-scalar truncation of the five-dimensional supergravity. As described above
$\mathcal{N}=1^*$ is a deformation of the ${\cal N}=4$ SYM superpotential by an
F-term, which is a complex operator. The gaugino bi-linear is also a complex
operator and, therefore, their sources are complex, making a total of 4 real
scalars in the dual supergravity. In addition, consistency of the 5$d$
supergravity theory requires that we include a $\textup{U}(1)$ gauge
field.\footnote{In ${\cal N}=1$ language, the $\textup{SU}(4)_R$ R-symmetry of
${\cal N}=4$ SYM decomposes as $\textup{SU}(3) \times \textup{U}(1)_R$. The
gauge field that we keep is dual to the $\textup{U}(1)_R$.} Altogether this
gives an $\textup{SO}(3)$ invariant truncation of $D=5$ supergravity that
involves the metric, 2 complex scalars, $\underline{m}=m e^{i \varphi}$ and
$\underline{\sigma}=\sigma e^{i \omega}$ and a $\textup{U}(1)$ gauge
field\footnote{This 4-scalar sector has been considered earlier
in~\cite{Pilch:2000fu, Bianchi:2000sm}. One can further truncate this theory by
setting either $\underline{m}=0$ or $\underline{\sigma}=0$. The
$\underline{m}=0$ truncation and its uplift to 10 dimensions has been discussed
in~\cite{Gubser:2009qm, Cassani:2010uw, Gauntlett:2010vu}. We discuss both cases
in Appendix~\ref{app:1-field}}. The GPPZ solution lives in a further
(consistent) truncation of this theory that keeps the norms $m$ and $\sigma$ and
truncates the angles $\varphi, \omega$ and the gauge field.

The uplift of the 4-scalar sector has a number of interesting features. It turns
out that the angles of the complex scalars are completely accounted for by a
combination of an $\textup{SO}(2)$ rotation of the coordinates of $S^5$ and an
$\textup{SO}(2)_{\textrm{IIB}}$ rotation of the $\textup{SL}(2)_{\textrm{IIB}}$
symmetry of IIB supergravity. To be more precise, let $(u^i, v^i)$ be
coordinates on $\mathbb{R}^6$ such that each triplet parameterises an
$\mathbb{R}^3$. The $S^5$ is then described by $u^2 + v^2=1$ and the
$\textup{SO}(3)$ symmetry acts by a simultaneous $\textup{SO}(3)$ rotation of
$u^i$ and $v^i$. The first $\textup{SO}(2)$ rotates the $u$'s into $v$'s and
geometrises the $\textup{U}(1)_R$ action of the dual QFT\@. The
$\textup{SO}(2)_{\textrm {IIB}}$ corresponds to the bonus $\textup{U}(1)$
symmetry of ${\cal N}=4$ SYM~\cite{intriligator:1998ig} and it is the
$\textup{U}(1)_Y$ group discussed in~\cite{Gunaydin:1984fk} in the context of
the $S^5$ compactification of IIB supergravity. It is also worth mentioning that
the periodicity of the angles $\varphi$ and $\omega$ in $D=5$ maps in ten
dimensions to the invariance under the combined operation of exchanging of $u^i$
with $v^i$ and performing an S-duality transformation.

Applying the uplift formulae to the GPPZ flow we find an explicit
ten-dimensional solution of type IIB, with all metric coefficients, the
dilaton-axion and the $p$-forms given in terms of elementary functions. All
$p$-forms are turned on in the solution, in particular both the NSNS and RR
2-form potentials (as mentioned in the previous paragraph, the solution is
invariant under $S$ duality, which exchanges the two potentials, followed by an
exchange of the coordinates $u^i$ and $v^i$). The ten-dimensional metric and the
axion-dilaton agree exactly with those of the partial uplift of
Pilch-Warner~\cite{Pilch:2000fu}. We checked using
Mathematica~\cite{Mathematica} that the type IIB equations are satisfied.

The solution is asymptotically $AdS_5 \times S^5$ and the leading correction is
due to the 2-form potentials, as expected based on the spectrum of linear
perturbations~\cite{Kim:1985ez} and the AdS/CFT dictionary. We also give the
first few sub-leading terms in the asymptotic expansion in order to compare with
other solutions that appeared in the literature. The first few sub-leading terms
are uniquely fixed by the boundary conditions~\cite{deHaro:2000vlm} and by
working out the asymptotic expansion to sufficiently high order one can extract
(in principle) the vevs of all gauge invariant operators using the method of
Kaluza-Klein holography~\cite{Skenderis:2006uy}. We will report on this
computation elsewhere.

The solution is still singular in ten dimensions but the divergence is milder
than that of the five-dimensional solution. In five dimensions the entire
spacetime metric goes to zero as we approach the singularity. In ten dimensions
the singularity structure depends on the size of the deformation parameter
relative to the gaugino condensate, which is quantified by a parameter
$\lambda$. It was argued in~\cite{Girardello:1999bd} that the singularity is
acceptable provided $\lambda \leq 1$. The singularity structure is different
depending on whether $\lambda <1$ or $\lambda=1$. The $\lambda=1$ is similar to
the $5d$ solution, with a singularity both in the non-compact directions and the
compact spherical part. When $\lambda <1$ the non-compact part of the metric is
now non-singular and there is only a singularity in the compact spherical part,
a singularity which was called ``ring singularity'' in~\cite{Pilch:2000fu}. More
precisely, following~\cite{Pilch:2000fu} we view $S^5$ as an $\mathbb{R}
\mathbb{P}^3$ fibered over a disc. The singularity is located at the edge of the
disc. In~\cite{Pilch:2000fu} it was argued that the singularity is associated
with 7-branes but we find no evidence for 7-branes in the near-singularity
structure of the metric.

It turns out that the singularity structure also depends on the choice of frame:
one can rescale the metric with appropriate powers of the scalars fields in the
solution. While in the usual frames, the Einstein and string frame, the
singularity structure is similar, there are also frames where the
ten-dimensional metric is only singular at one point, a point at the edge of the
disc. It would be interesting to understand the physics behind this observation.

It is possible that the supergravity approximation is not sufficient to describe
duals of $\mathcal{N} = 1^*$ theories and that some stringy mechanism is needed
to resolve the singularity. In~\cite{Polchinski:2000uf} Polchinski and Strassler
suggested that the five-dimensional singularity is resolved by D3-branes
polarised via Myers' effect~\cite{Myers:1999ps} into five-branes with
world-volume $\mathbb{R}^4 \times S^2$, where $S^2$ is an equator of $S^5$ and
$\mathbb{R}^4$ is a slice of $AdS_5$ at fixed radius. The construction is
motivated by the observation that the F-term condition~\eqref{Ftermd} tell us
that the branes are non-commutatively expanded into two spheres.\footnote{The
scalar components of the superfields $\Phi_i$ are the transverse coordinates of
the branes ($x^i = 2 \pi \alpha^\prime \phi_i$).} Consider, for instance, the
Higgs vacuum. The Dirac-Born-Infeld-Wess-Zumino action for the D3 branes
becomes~\cite{Myers:1999ps}
\begin{eqnarray}
S \sim \mu_3 \int C_4 + \mu_3 (2 \pi \alpha^\prime)^2 \int F_{0123ijk} [ x^i, x^j] x^k
\;,
\end{eqnarray}
and we see that the expanded D3-branes have an additional electric coupling to
the RR 6-form and are therefore equivalent to a single D5-brane with $N$ units
of D3-brane charge and zero net D5-brane charge. Then the Higgs vacuum is
identified with a single D5-brane. S-duality transformations can be used to
construct the backgrounds dual to the other vacua. The confining vacua are
identified with single NS-branes, while the others are superpositions of D5 and
NS5 branes. This interpretation also agrees with the AdS/CFT dictionary since
the fermionic mass term in~\eqref{eq:massterm} corresponds the lowest KK-mode in
the expansion of the complex two-form field of type IIB and this latter is
exactly the potential needed to polarise D3-branes into 5-branes. However, only
the asymptotic solutions near the boundary and the branes were
given~\cite{Polchinski:2000uf} and the full supergravity solutions corresponding
to the various brane configurations are not known. Comparing this solution with
the uplifted GPPZ solution, we find that the boundary conditions (the sources)
agree, but the subleading terms in the near-boundary expansion disagree. Since
the first few sub-leading terms are uniquely fixed in terms of the sources, we
conclude that the Polchinski-Strassler $10d$ field do not satisfy the IIB
equations near the boundary.

The finite temperature physics of $\mathcal{N}=1^*$ theory was studied
holographically by Freedman and Minahan~\cite{Freedman:2000xb} by deforming the
solution of a black D3-brane by adding the complex combination of the RR and NS
3-forms dual to the mass terms. The solution is obtained perturbatively up to
second order in the deformation parameter. We can use $T=0$ limit of
Freedman-Minahan solution as a perturbative check of our solution in the UV\@,
and we find that the two solutions are indeed in agreement.

The paper is organised as follows. In the next section we review the
five-dimensional GPPZ solution. In Section~\ref{sec:uplift_gppz} we present the
uplift of the 4 scalar sector of the $D=5$ supergravity and of the GPPZ
solution. In Section~\ref{sec:asymptotics} we discuss the near-boundary
asymptotics and in Section~\ref{sec:singularity} we analyse the singularity
structure of the ten-dimensional solution. We conclude with an outlook in
Section~\ref{sec:conclusion}. The paper contains also a number of appendices. In
Appendix~\ref{app:IIb_uplift} we summarise the uplift formulas
from~\cite{Baguet:2015sma} and in Appendix~\ref{app:E6} we present the explicit
parameterisation of the scalar E$_{6(6)}$ matrix in the 4-scalar truncation of
$D=5$ maximal supergravity. In Appendix~\ref{app:uplift_in_pw_coords} we present
the ten-dimensional solution in Pilch-Warner coordinates and in
Appendix~\ref{app:harmonics} we list the $\textup{SO}(3)$ invariant spherical
harmonics that we use in the main text. Finally, in Appendix~\ref{app:1-field}
we present the uplift of the two one-scalar truncations of the GPPZ solution.

%%%%%%%%%%%%%%%%%%%%%%%%%%%%%%%%%%%%%%%%%%%%%%%%%%%%
%
% GPPZ
%
%%%%%%%%%%%%%%%%%%%%%%%%%%%%%%%%%%%%%%%%%%%%%%%%%%%%
\section{The GPPZ flow}%
\label{sec:GPPZ}

The GPPZ flow~\cite{Girardello:1999bd} is a solution of the five-dimensional
$\mathcal{N}=8$ $\textup{SO}(6)$ gauged supergravity~\cite{Pernici:1985ju,
Gunaydin:1984qu}. The field content of the theory can be organised in
representations of the $\textup{SO}(6) \times \textup{SL}(2)$ subgroup of
$\textup{E}_{6(6)}$. It consists of 15 massless vectors fields in the adjoint of
$\textup{SO}(6)$, 12 topologically massive two-forms transforming in the $({\bf
6}, {\bf 2})$ of $\textup{SO}(6) \times \textup{SL}(2)$ and 42 scalars
parameterising an $\textup{E}_{6(6)}/\textup{USp}(8)$ coset space and
transforming as
\begin{eqnarray}
\label{scalarrep}
{\bf 42} = {\bf 20^\prime}_{(0)} + {\bf 10}_{(-2)} + {\bf \overline{10}}_{(2)} + {\bf 1}_{(4)} + {\bf 1}_{(-4)} \, ,
\end{eqnarray}
where the subscripts are the charges under the $\textup{U}(1)_Y$ subgroup of
$\textup{SL}(2)$~\cite{Gunaydin:1984fk}. The masses of these scalar are $m^2 = -
4$ (for the ${\bf 20^\prime}$), $m^2=-3$ (for the ${\bf 10}$ and ${\bf
\overline{10}}$) and $m^2=0$ (for the two ${\bf 1}$s).

According to the AdS/CFT dictionary, the 42 scalars are dual to relevant and
marginal operators\footnote{Recall that the mass of an AdS scalar is mapped to
the conformal dimension of the dual field theory operator by $\Delta = 2 +
\sqrt{4 + m^2}$.} of $\mathcal{N}=4$ SYM\@. $\mathcal{N}=4$ SYM contains six
scalars $\phi_i$ and four fermions $\lambda_a$ transforming in the ${\bf 6}$ and
${\bf 4}$ of $\textup{SU}(4)$, respectively. Then the scalars in the $ {\bf
20^\prime}$ correspond to scalar bilinears ${\rm Tr}\phi_{(i} \phi_{j)}
(-$traces) of conformal dimension $\Delta =2$, the scalars in the $ {\bf 10}$
are massive deformations with $\Delta =3$, schematically \begin{eqnarray}
\label{10r} Q^2 {\rm Tr}\phi_{(i} \phi_{j)} \sim {\rm Tr}(\lambda_a \lambda_b +
\phi^3) \end{eqnarray} and the scalars in the $ {\bf 1}$ are the gauge-coupling
deformations.

We are interested in massive deformations of $\mathcal{N}=4$ $\textup{SU}(N)$
SYM that break supersymmetry to $\mathcal{N}=1$. In $\mathcal{N}=1$ notation the
six scalar fields $\phi_i$ are arranged in three complex scalars, which together
with three of the four fermions form three chiral superfields $\Phi_i $,
$i=1,2,3$, while the remaining fermion sits in the vector multiplet. Of the full
$\textup{SO}(6)$ R-symmetry only a $\textup{U}(1)_R \times \textup{SU}(3)_R$
subgroup is manifest, under which the vector superfield is neutral and the three
chiral superfields transform in the ${\bf 3}_{2/3}$.

To identify the relevant 5$d$ scalars we need to describe the mass deformation
in more detail. In ${\cal N}=1$ language the mass deformation modifies the
superpotential of the theory by the addition of the term
\begin{equation}%
\label{eq:masstermbis}
\delta \mathcal{W} = m_{ij} {\rm tr} (\Phi_i \Phi_j)
\end{equation}
that only involves three of the four $\mathcal{N}=4$ fermions. Thus we need to
decompose the $\textup{SU}(4)\sim \textup{SO}(6)$ representations into
$\textup{SU}(3) \times \textup{U}(1)$ and single out the gaugino. The
fundamental of $\textup{SU}(4)$ splits as
\begin{equation}%
\label{4split}
{\bf 4} \rightarrow {\bf 3} + {\bf 1}
\end{equation}
and thus the four fermions $\lambda_a$ in the ${\bf 4}$ of $\textup{SU}(4)$
split into the ${\bf 3}$ corresponding to the three fermions in the chiral
multiplets $\Phi_i$ and the {\bf 1} which is the gaugino $\lambda$. The
fermionic mass term in the ${\bf 10}$ then decomposes as
\begin{eqnarray}
 \label{10split}
{\bf 10} \to {\bf 1} + {\bf 3} + {\bf 6} \, ,
\end{eqnarray}
and we identify the ${\bf 6}$ with the mass deformation, while the scalar in the
${\bf 1}$ corresponds to a gaugino condensate. Integrating out the auxiliary
fields we find that ${\cal N}=4$ SYM the generic mass deformation also involves,
on the gravity side, part of the scalars in the ${\bf 20^\prime}$,\footnote{The
mass deformation also involves the Konishi operator, which however decouples in
the supergravity limit.} unless the three fermion masses are taken to be equal
(the details will be reported in~\cite{toappear}). If the masses are equal, the
part corresponding to the ${\bf 20^\prime}$ does not appear, and there is a
residual $\textup{SO}(3)$ symmetry that allows us to keep only two holomorphic
scalars, $\underline{\sigma} \in {\bf 1}$ and $\underline{m} \in {\bf 6}$, while
setting all the remaining fields consistently to zero. These two fields are dual
to the operators\footnote{These operators are obtained from ${\bf 20^\prime}$ by
acting with two supercharges and they contain also a part proportional to
$\phi^3$ that we suppress here.}
\begin{equation}
\mathcal{O}_3 = \sum_{i=1}^3 {\rm tr} (\lambda_i \lambda_i),    \qquad \mathcal{O}_4 =  {\rm tr} (\lambda_4 \lambda_4).
\end{equation}
Similarly, we get two anti-holomorphic scalars, $\bar{\underline{\sigma}},
\bar{\underline{m}}$ (the complex conjugates of $\underline{\sigma}$ and
$\underline{m}$) from the ${\bf \overline{10}}$. This is as expected, since
$\underline{m}$ and $\underline{\sigma}$ are dual to chiral operators.

In AdS/CFT the QFT generating functional of correlation functions becomes the
on-shell value of the bulk action. Since $\underline{m}$ and
$\underline{\sigma}$ couple to complex operators, the generating functional will
only contain the modulus of $\underline{m}$ and $\underline{\sigma}$. Indeed, in
${\cal N}=4$ SYM, $\langle \mathcal{O}_3 \mathcal{O}_3 \rangle = \langle
\overline{\mathcal{O}}_3 \overline{\mathcal{O}}_3\rangle =0$ but $\langle
\mathcal{O}_3 \overline{\mathcal{O}}_3\rangle \neq 0$ and the same with
$\mathcal{O}_3 \to \mathcal{O}_4$, which means the generating function will
depend on $|\underline{m}|^2$, $|\underline{\sigma}|^2$ but not on
$\underline{m}^2,  \bar{\underline{m}}^2, \underline{\sigma}^2,
\bar{\underline{\sigma}}^2$, and similarly for the contributions coming from
higher point functions. Indeed, we will see in the next section that there is
consistent truncation of the bulk supergravity to the moduli $m$ and $\sigma$ of
$\underline{m}$ and $\underline{\sigma}$,
\begin{eqnarray}
\underline{m}=m\,e^{i \varphi}\;,\qquad
\underline{\sigma}=\sigma\,e^{i \omega} \, .
%\label{angles}
\end{eqnarray}

We are thus lead to look for 5$d$ solutions of the form
\begin{eqnarray}
\label{5dmets}
d s^2 = d y^2 + e^{2 \phi(y)} d x^\mu d x_\mu
\end{eqnarray}
with $\mu=0, \dots, 3$ and non-trivial profile for the real fields $m(y)$ and
$\sigma(y)$. The radial coordinate $y$ ranges from $- \infty$ (IR) and $+
\infty$ (UV). With this truncation the Lagrangian reduces to
 \begin{eqnarray}
 \label{2rslag}
L &=& \sqrt{-g} \left\{ - \frac{1}{4} R + \frac{1}{2} (\partial m)^2 + \frac{1}{2} (\partial \sigma)^2 \right. \nonumber \\
& & \left. - \frac{3}{8} \left[ (\cosh \frac{2 m}{\sqrt{3}})^2 + 4 \cosh \frac{2 m}{\sqrt{3}} \cosh 2 \sigma - (\cosh 2 \sigma )^2 + 4 \right] \right\} \, .
\end{eqnarray}
Because of supersymmetry, the fields $\phi$, $m$ and $\sigma$ satisfy the first
order equations
\begin{eqnarray}
\label{5deqm}
&& \dot{\phi} = \frac{1}{2} \left[ \cosh \frac{2 m}{ \sqrt{3}} + \cosh 2 \sigma\right] \, , \nonumber \\
&&\dot{m} = - \frac{\sqrt{3}}{2} \sinh \frac{2 m}{ \sqrt{3}} \, , \nonumber \\
&& \dot{\sigma} = - \frac{3}{2} \sinh 2 \sigma \, ,
\end{eqnarray}
descending form the superpotential
\begin{eqnarray}
\label{5superp}
W = \frac{3}{4} \left[ \cosh \frac{2 m}{ \sqrt{3}} + \cosh 2 \sigma \right] \, .
\end{eqnarray}
The solution, which is often denoted as GPPZ flow~\cite{Girardello:1999bd}, is
\begin{align}
\label{5rsol}
m(y) &= \frac{\sqrt{3}}{2} \log \left[ \frac{1 + e^{-(y - C_1)}}{1 - e^{-(y - C_1)}} \right] = \sqrt3\operatorname{arctanh} e^{-(y-C_1)} \, , \nonumber \\
\sigma(y) &= \frac{1}{2} \log \left[ \frac{1 + e^{-3(y - C_2)}}{1 - e^{-3(y - C_2)}} \right] = \operatorname{arctanh} e^{-3(y-C_2)} \, , \nonumber \\
\begin{split}
\phi(y) &= y + \frac12\log\left[1-e^{-2(y-C_1)}\right] + \frac16\log\left[1-e^{-6(y-C_2)}\right] \\
&= y - \log\cosh\frac{m(y)}{\sqrt{3}} - \frac13\log\cosh\sigma(y) \, .
\end{split}
\end{align}
where $C_1$ and $C_2$ are two arbitrary integration constants.\footnote{The
integration constants $C_i$ used here are identical to those used by Pilch \&
Warner~\cite{Pilch:2000fu}, and are related to those used by
GPPZ~\cite{Girardello:1999bd} by $C_1^\text{(GPPZ)} = C_1$, $C_2^\text{(GPPZ)} =
3C_2$. Also the definition of $\phi(y)$ differs between Pilch \& Warner and
GPPZ\@. Here we are using the Pilch \& Warner definition, which is related to
GPPZ by $\phi^{(GPPZ)} = \phi - (C_1 + C_2)/2$.}

Generically, solutions of the type~\eqref{5rsol} can represent both deformations
of the dual field theory by an operator $\mathcal{O}$ and/or different vacua of
the same theory characterised by a vev $ \langle \mathcal{O} \rangle$. The
behaviour of the solution in the asymptotic AdS region, $y \to + \infty$,
discriminates between the two options. For $y \to + \infty$, the asymptotic
behaviour consists of a non-normalisable part and a normalisable one
\begin{equation}
\varphi \underset{y \to + \infty} \sim  e^{(\Delta- 4)y}(A  + \cdots) + e^{ - \Delta y} (B + \cdots) \, ,
\end{equation}
where $\Delta$ is the conformal dimension of the dual operator and the dots in
the leading non-normalizable part are local functions of $A$ while the dots in
the normalizable part are functions of both $A$ and $B$. The coefficient $A$ of
the non-normalisable solution is interpreted as a deformation of the Lagrangian
while the coefficient $B$ of the normalisable solution is related to the vev
$B=1/(2 \Delta -4) \langle O \rangle$, where $O$ is the operator dual to
$\varphi$~\cite{deHaro:2000vlm}.

For $y\to +\infty$, the GPPZ solution behaves as
\begin{eqnarray}
\label{eq:gppz_asymptotics}
&& \phi(y) \underset{y \to + \infty} \sim y \\
&&  m(y)  \underset{y \to + \infty} \sim  m_0 e^{-y}, \qquad m_0 = \sqrt{3}e^{C_1} \nonumber\\
&&  \sigma(y) \underset{y \to + \infty} \sim  \frac{1}{2} \sigma_0 e^{ - 3 y}, \qquad \sigma_0 = 2 e^{3 C_2}  \, . \nonumber
%2\sigma_0 e^{ - 3 y}, \qquad \sigma_0 = \frac12 e^{C_2}  \, . \nonumber
\end{eqnarray}
From these asymptotics we see that, since $\Delta =3$, $m_0$ corresponds to a
mass deformation and $\sigma_0 = {\rm Re} \langle \lambda\lambda \rangle$ is the
real part of the gaugino condensate. It is then natural to interpret the
solution as a flow from the mass deformed $\mathcal{N}=4$ to $\mathcal{N}=1^*$
in the IR\@.

The metric has a naked singularity for $y \to C_1$ (with $y \geq C_1$),
\begin{eqnarray}
d s^2 = d y^2 + a(y - C_1) d x^\mu d x_\mu + \dots,
\end{eqnarray}
where $a = 2 e^{C_1 + C_2} (2 \sinh \left(3 (C_1-C_2)\right))^{1/3}$. The Ricci
scalar is singular
\begin{equation}
R = -(y-C_1)^{-2} + \dots
\end{equation}
and there is no change of frame in which the singularity disappears or is
milder. Notice that also the solution for $m$ diverges at $y=C_1$
\begin{eqnarray}
 m(y) = - \frac{\sqrt{3}}{2} \log (y - C_1) + \dots
\end{eqnarray}
while the behaviour of $\sigma$ depends on the relation between $C_1$ and $C_2$.
If $C_2 \leq C_1$ then $\sigma$ is regular.

Singularities of this kind are common in most 5$d$ solutions and criteria have
been proposed to establish whether the solutions are physically acceptable or
not. In particular in~\cite{Gubser:2000nd} it was proposed that a singular
solution is physically acceptable if it can be obtained as the zero temperature
limit of a regular black-hole. The conditions for the existence of the black
hole solution constrain the parameters of the singular solution. In this case
the criterion gives $C_2 \leq C_1$. By looking at the behaviour of Wilson loops
it was shown in~\cite{Girardello:1999bd} that the solutions with $C_2 \leq C_1$
confines. Such solutions should then be dual to the confining vacua of
$\mathcal{N}=1^*$.

%%%%%%%%%%%%%%%%%%%%%%%%%%%%%%%%%%%%%%%%%%%%%%%%%%%%
%
% Uplift of the GPPZ solution
%
%%%%%%%%%%%%%%%%%%%%%%%%%%%%%%%%%%%%%%%%%%%%%%%%%%%%
\section{Uplift of the GPPZ solution}%
\label{sec:uplift_gppz}

The general uplift of $\mathcal{N}=8$ $\textup{SO}(6)$ gauged supergravity to type IIB
was constructed in~\cite{Baguet:2015sma} and we recall the main formulae in
Appendix~\ref{app:IIb_uplift}. In this section we first review the 4-scalar
truncation of $\mathcal{N}=8$ supergravity in which the GPPZ solution lives and
then we apply to it the uplift formulae of Appendix~\ref{app:IIb_uplift} %from
the last section in this 4-scalar truncation to obtain the full IIB uplift of
the GPPZ solution. Finally, we explicitly verify the entire set of IIB field
equations is satisfied by the ten-dimensional solution.

\subsection{Four-scalar truncation of \texorpdfstring{$D=5$}{D=5} supergravity}

As discussed above, an important ingredient in the construction of the GPPZ
solution is the invariance under an $\textup{SO}(3)$ subgroup of the gauge group
$\textup{SO}(6)$ that allows to truncate the full theory to a pair of complex
scalars~\cite{Girardello:1999bd}. Even if the GPPZ solution was found in a
truncation involving only two real scalars, one can actually embed the flow in a
larger theory that is obtained by truncating the $\mathcal{N}=8$ supergravity to
the full set of $\textup{SO}(3)$ invariant fields~\cite{Pilch:2000fu}. This
gives an $\mathcal{N}=2$ supergravity coupled to two hyper-multiplets. Of the 42
scalars~\eqref{scalarrep} of the $\mathcal{N}=8$ theory we only keep the 8
singlets under the
\begin{eqnarray}
\textup{SO}(3)_{\rm diag}\subset\textup{SO}(3)\times \textup{SO}(3)\subset\textup{SO}(6)
\label{SO3}
\end{eqnarray}
subgroup of the gauge group $\textup{SO}(6) \sim \textup{SU}(4)$. These form the coset space
$\textup{G}_{2(2)}/\textup{SO}(4)$ and are dual to the operators
\begin{align}
& \mathcal{O}_1 = \sum_{i =1}^3 \left( {\rm tr} (\phi_i \phi_i) - {\rm tr} (\phi_{i+3} \phi_{i+3}) \right),   & \mathcal{O}_2 &= \sum_{i=1}^3 {\rm tr} (\phi_i \phi_{i+3}), \nonumber \\
&\mathcal{O}_3 = \sum_{i=1}^3 {\rm tr} (\lambda_i \lambda_i)    &\mathcal{O}_4 & =  {\rm tr} (\lambda_4 \lambda_4) \\
& \mathcal{O}_5 = (F^+)^2, &\mathcal{O}_6 &= (F^-)^2, \nonumber
\end{align}
where $F^\pm$ is the (anti)-self-dual field gauge strength. $\mathcal{O}_1$ and
$\mathcal{O}_2$ are the $\textup{SO}(3)_{\rm diag}$ singlets contained in ${\bf
20}^\prime$, the complex operators $\mathcal{O}_3$ and $\mathcal{O}_4$ are the
$\textup{SO}(3)_{\rm diag}$ singlets in the ${\bf 10}$ and ${\bf
\overline{10}}$, and $\mathcal{O}_5, \mathcal{O}_6$ correspond to the two
singlets. Among the $\textup{SU}(4)_R$ gauge fields, the truncation to singlets
under~\eqref{SO3} only keeps a single $\textup{U}(1)$ gauge field, dual to the
$\textup{U}(1)_R$ subgroup of $\textup{SU}(4)_R \to \textup{SU}(3) \times
\textup{U}(1)_R$.\footnote{We normalize the $\textup{U}(1)$ such that the
charges are those of the QFT, see the discussion around (6.88)
in~\cite{Bianchi:2001kw}.}

The further truncation to the 2 complex scalars dual to $\mathcal{O}_3$ and
$\mathcal{O}_4$ can also be shown to be consistent as it corresponds to the
truncation to singlets under an additional discrete subgroup within
$\textup{U}(1)_R \times \textup{U}(1)_Y$, see~\cite{Pilch:2000fu} for details.
In field theory the discrete $\textup{U}(1)_R$ transformation is
\begin{equation}
(\phi_i, \phi_{i+3}) \to (\phi_{i+3}, -\phi_{i}), \qquad
\lambda_i \to e^{-\frac{\pi}{4}i } \lambda_i, \qquad \lambda_4 \to e^{\frac{3 \pi}{4} i} \lambda_4
\end{equation}
Thus under this transformation
\begin{equation} \label{U(1)Rdiscrete}
 \mathcal{O}_1 \to - \mathcal{O}_1, \quad \mathcal{O}_2 \to -\mathcal{O}_2, \quad \mathcal{O}_3 \to e^{-\frac{\pi}{2}i } \mathcal{O}_3, \qquad \mathcal{O}_4 \to e^{-\frac{\pi}{2}i } \mathcal{O}_4
 %\quad \mathcal{O}_5 \to \mathcal{O}_5, \quad \mathcal{O}_6 \to \mathcal{O}_6
\end{equation}
while $\mathcal{O}_5$ and $\mathcal{O}_6$ are invariant. Under $\textup{U}(1)_Y$
the operators $\mathcal{O}_1$ and $\mathcal{O}_2$ are neutral, $\mathcal{O}_3$
and $\mathcal{O}_4$ have charge $2$ (and the complex conjugates charge $-2$) and
$\mathcal{O}_5$ and $\mathcal{O}_6$ have charges $4$ and $-4$
(see~\eqref{scalarrep}). Thus the combined operation of the discrete
$\textup{U}(1)_R$ in~\eqref{U(1)Rdiscrete} and an discrete $\textup{U}(1)_Y$
rotation with angle $\pi/4$ yields
\begin{equation}
 \mathcal{O}_1 \to - \mathcal{O}_1, \quad \mathcal{O}_2 \to -\mathcal{O}_2, \quad \mathcal{O}_3 \to \mathcal{O}_3, \qquad \mathcal{O}_4 \to \mathcal{O}_4,
 \quad \mathcal{O}_5 \to - \mathcal{O}_5, \quad \mathcal{O}_6 \to -\mathcal{O}_6
\end{equation}
thus projecting out $\mathcal{O}_1, \mathcal{O}_2, \mathcal{O}_5, \mathcal{O}_6$
and keeping $\mathcal{O}_3$ and $\mathcal{O}_4$.

We parameterise the scalars dual to $\mathcal{O}_3$ and $\mathcal{O}_4$ as
\begin{eqnarray}
\underline{m}=m\,e^{i \varphi}\;,\qquad
\underline{\sigma}=\sigma\,e^{i \omega} \, .
\label{angles}
\end{eqnarray}
The five-dimensional theory~\cite{Gunaydin:1985cu} then reduces to
\begin{eqnarray}
\frac{1}{\sqrt{g}}\, {\cal L} &=&
-\frac14\,R
-\frac{1}{12}\,F_{\mu\nu}\,F^{\mu\nu}
-\frac{1}{54}\,\epsilon^{\mu\nu\rho\sigma\tau}\,
A_\mu F_{\nu\rho} F_{\sigma\tau}
+ \frac12\,\partial_\mu m\,\partial^\mu m
+ \frac12\,\partial_\mu \sigma\,\partial^\mu \sigma
\nonumber\\
&&{}
+\frac38\,\sinh^2\left(\frac{2m}{\sqrt{3}}\right)D_\mu\varphi D^\mu\varphi
+\frac18\,\sinh^2\left(2\sigma\right)D_\mu\omega D^\mu\omega
-V_{\rm pot}
  \;,\quad
  \label{D5_4scalars}
\end{eqnarray}
with the Maxwell and Chern-Simons terms of minimal supergravity, covariant
derivatives
\begin{eqnarray}
D_\mu\omega&=& \partial_\mu \omega + 2\,A_\mu\;,\qquad
D_\mu\varphi~=~\partial_\mu \varphi + \frac23\,A_\mu
\;,
\label{cov}
\end{eqnarray}
and the scalar potential
\begin{eqnarray}
V_{\rm pot} &=&-
\frac{3}{8} \left(4 \cosh \left(\frac{2 m}{\sqrt{3}}\right) \cosh (2 \sigma )+\cosh
  ^2\left(\frac{2 m}{\sqrt{3}}\right)-\cosh ^2(2 \sigma )+4\right)
  \;,
  \label{D5potential}
\end{eqnarray}
which only depends on the absolute values of the complex scalars. The scalar
kinetic term is an $\left(\textup{SU}(1,1)/\textup{U}(1)\right)^2$ coset space,
and the covariant derivatives~\eqref{cov} correspond to the gauging of
$\textup{U}(1)_R$

Note that the angles $\varphi, \omega$ source the Maxwell equation
\begin{eqnarray}
\nabla_\nu F^{\mu\nu}
+\frac{1}{6}\,\epsilon^{\mu\nu\rho\sigma\tau}\,F_{\nu\rho}F_{\sigma\tau}
&=&
\frac32\,\sinh^2\left(\frac{2m}{\sqrt{3}}\right)D_\mu\varphi+\frac32\,
\sinh^2\left(2\sigma\right)D_\mu\omega
\nonumber\\
&\equiv&
\frac32\,{\cal J}^\mu
\;.
\label{5DMaxwell5}
\end{eqnarray}
Thus one may either set the vector field to zero and consider constant angles or
demand that the angles are covariantly constant,
\begin{equation}
D_\mu\varphi=D_\mu\omega=0
\end{equation}
This condition is solved by
\begin{eqnarray} \label{lambda-source}
\omega=-\lambda\;,\qquad
\varphi=-\frac13\,\lambda\;,\qquad
A_\mu = \frac12\,\partial_\mu \lambda
\;,
\end{eqnarray}
for any spacetime dependent function $\lambda(x)$.

With the angles being (covariantly) constant, the field equations for $m$ and
$\sigma$ decouple and the GPPZ flow~\eqref{5rsol} is still a solution. In the
uplift formulae we will also employ the variables~\cite{Pilch:2000fu}
\begin{eqnarray}\label{eq:def_mu_nu}
\mu&\equiv& e^{\sigma}\;,\qquad
\nu~=~ e^{m/\sqrt{3}}\;,
\end{eqnarray}
in terms of which the flow equations~\eqref{5deqm} take the form
\begin{eqnarray}
\dot{\mu} &=& \frac3{4\mu}\,(1-\mu^4)
\;, \nonumber\\
\dot{\nu} &=& \frac1{4\nu}\,(1-\nu^4)
\;,\nonumber\\
\dot{\phi} &=& \frac1{4\mu^2}\,(1+\mu^4)+\frac1{4\nu^2}\,(1+\nu^4)
\;.
\label{flowmunu}
\end{eqnarray}

\subsection{Uplift of the 4-scalar truncation: metric and dilaton/axion}

In order to apply the explicit uplift formulae given in
Appendix~\ref{app:IIb_uplift}, we first evaluate the matrix~\eqref{MD5} for the
four-scalar truncation $(\mu, \nu, \varphi, \omega)$ by exponentiating the
associated generators in the group $\textup{E}_{6(6)}$\,. We give some details
in Appendix~\ref{app:E6}. Since all scalars are singlets of the $\textup{SO}(3)$
in~\eqref{SO3} it proves useful to decompose the $S^5$ sphere harmonics ${\cal
Y}^a$ into
\begin{eqnarray}
{\cal Y}^a &\longrightarrow&
\{u^i, v^i\}
\;,
\end{eqnarray}
with $u^i u^i+v^i v^i=1$\,. Moreover, for compactness of notation, it is useful
to define the rotated functions
\begin{eqnarray}
U^i &=& \cos\left(\frac14(\varphi+\omega)\right)\,u^i + \sin\left(\frac14(\varphi+\omega)\right)\,v^i
\;,\nonumber\\
V^i &=& \cos\left(\frac14(\varphi+\omega)\right)\,v^i - \sin\left(\frac14(\varphi+\omega)\right)\,u^i
\;,
\label{UVrotated}
\end{eqnarray}
where $\varphi$ and $\omega$ are the $x$-dependent phases of the scalars $m$ and
$\sigma$ of the $D=5$ theory, see~\ref{angles}. This transformation is a local
$\textup{U}(1)$ corresponding to the $\textup{U}(1)_R$ of the dual QFT\@.
Similarly, we define the rotated one-forms
\begin{eqnarray}
\Theta^i &=& \cos\left(\frac14(\varphi+\omega)\right)\,Du^i + \sin\left(\frac14(\varphi+\omega)\right)\,Dv^i
\;,\nonumber\\
\Lambda^i &=& \cos\left(\frac14(\varphi+\omega)\right)\,Dv^i - \sin\left(\frac14(\varphi+\omega)\right)\,Du^i
\;, \label{ThetaLambda}
\end{eqnarray}
where the covariant derivatives
\begin{eqnarray}
Du^i&\equiv&du^i-\frac{1}{3}\,v^i A_\mu\,dx^\mu\;,\nonumber\\
Dv^i&\equiv&dv^i+\frac{1}{3}\,u^i A_\mu\,dx^\mu
\;,
\end{eqnarray}
correspond to the Kaluza-Klein basis~\eqref{DY}. These are the objects that
naturally appear in the uplift formulae of Appendix~\ref{app:IIb_uplift}. Let us
also note that the proper identification of the $\textup{U}(1)$ vector field
$A_\mu$ among the 15 $\textup{SO}(6)$ fields $A_\mu{}^{ab}$ gives rise to the
relations
\begin{eqnarray}
F_{\mu\nu}{}^{ab} \, M_{ab,cd} \, F^{\mu\nu\,cd}
&=& \frac43\,F_{\mu\nu} F^{\mu\nu}
\;,
\nonumber\\
\varepsilon_{abcdef}\,F_{\mu\nu}^{ab} F_{\rho\sigma}^{cd} A_\tau^{ef} &=&
-\frac{32\sqrt{2}}{9}\,F_{\mu\nu} F_{\rho\sigma} A_\tau
\;.
\label{vec_rel}
\end{eqnarray}

We can now give the fields of the uplifted solution. The IIB
metric~\eqref{metricIIB} takes the explicit form
\begin{eqnarray}
ds_{\rm IIB}^2 &=&
 \Delta^{-2/3}\left({g}_{\mu\nu}(x) \, dx^\mu dx^\nu
 + \Delta^{8/3}\, d\hat{s}_5^2 \right)
 \label{metricIIBexplicit}
\end{eqnarray}
with the warp factor $\Delta$ and the internal metric $d\hat{s}_5^2$ given by
\begin{eqnarray}
\Delta^{-8/3} &=&
\frac{ \left(1+\mu ^2 \nu ^2\right)^3 \left(\mu ^2+\nu ^6\right)}{16\, \mu ^4
  \nu ^6}
  + \frac{U^2\,V^2}{16\, \mu ^4 \nu ^8}\,
  \left(1-\nu ^4\right)^2 \left(\mu ^2-\nu ^2\right)^2 \left(1+\mu ^2 \nu^2\right)^2
  \nonumber\\
  &&{}
  -\frac{(U\cdot V)^2}{16\, \mu ^4 \nu ^8}\,
  \left(1-\nu ^4\right)^2 (1-\mu^2 \nu^2 )^2 \left(\mu
  ^2+\nu ^2\right)^2
  \;,
  \label{warp}
\\[2ex]
d\hat{s}_5^2 &=&
\frac{ \left(1+\nu ^4\right) \left(\mu ^2+\nu ^2\right) \left(1+\mu ^2 \nu
  ^2\right)}{8\, \mu ^2 \nu ^4}
 \left( \Theta^i\Theta^i + \Lambda^i\Lambda^i \right)
\nonumber\\
&&{}
  -\frac{ \left(1-\nu ^4\right)^2}{8\, \nu ^4}
  \left((U^2-V^2)\,(\Theta^i\Theta^i - \Lambda^i\Lambda^i)+4\,(U\cdot V) \,\Theta^i\Lambda^i \right)
  \nonumber\\
  &&{}
  +\frac{ \left(1-\mu ^4\right) \left(1-\nu
  ^4\right)}{8\, \mu ^2 \nu ^2}
  \left((U^2-V^2)\,(\Theta^i\Theta^i - \Lambda^i\Lambda^i)-4\,(U\cdot V) \,\Theta^i\Lambda^i \right)
\nonumber\\
&&{}
+\frac{\left(1-\mu^4\nu^4\right) (1-\mu^2 \nu^2)
\left(\mu^2+\nu ^6\right)}{16\, \mu ^4 \nu ^6}
 \left((V^i\Theta^i)(V^j\Theta^j)  + (U^i\Lambda^i )(U^j \Lambda^j)\right)
\nonumber\\
&&{}
   + \frac{ \left(1-\mu ^4 \nu ^4\right) \left(1+\mu ^2 \nu ^2\right)
   \left(\mu^2-\nu ^6\right)}{8\, \mu ^4 \nu ^6}\, (V^i \Theta^i)(U^j \Lambda^j)
\nonumber\\
&&{}
  -\frac{(\mu^4-\nu^4)(1-\nu^8)}{4\,\mu^2\nu^6} \, (U^i \Theta^i) (V^j \Lambda^j)
  \;.
  \label{G5}
\end{eqnarray}
For vanishing angles (i.e.\ $U^i=u^i$, $V^i=v^i$, $\Theta^i=du^i$,
$\Lambda^i=dv^i$) we recover the result from~\cite{Pilch:2000fu}.\footnote{We
corrected a typo in~\cite{Pilch:2000fu} in the form of index contractions of the
penultimate term in~\eqref{G5}.} It is important to note that the only
singularities of the IIB metric can be located at $\mu,\nu=0$ or
$\mu,\nu=\infty$. Indeed, the warp factor~\eqref{warp} can be estimated (using
that $U^2\,V^2\ge(U\cdot V)^2$, and $U^2\,V^2=U^2\,(1-U^2)\le\frac14$) to be
\begin{eqnarray}
\Delta^{-8/3} &\ge&
  \frac{ \left(1+\mu ^2 \nu ^2\right)^3 \left(\mu ^2+\nu ^6\right)}{16\, \mu ^4
  \nu ^6}
  - \frac{U^2\,V^2}{4\, \mu ^2 \nu ^6}\,
  \left(1-\mu ^4\right) \left(1-\nu ^4\right)^3
  \nonumber\\
  &\ge&
  \frac{ \left(1+\mu ^2 \nu ^2\right)^3 \left(\mu ^2+\nu ^6\right)}{16\, \mu ^4
  \nu ^6}
  - \frac{1}{16\, \mu ^2 \nu ^6}\,
  \left(1-\mu ^4\right) \left(1-\nu ^4\right)^3
   \nonumber\\
  &=&
 \frac{ \left(\mu ^2 + \nu ^2\right)^3 \left(1+\mu ^2\nu ^6\right)}{16\, \mu ^4  \nu ^6} ~>~0\;.
\end{eqnarray}
We will take a closer look at the possible singularities in
Section~\ref{sec:singularity}.

For the symmetric $\textup{SL}(2)$ dilaton/axion matrix $m_{\alpha\beta}$,
\begin{eqnarray}
m_{\alpha\beta}&=&
\frac{1}{\Im \tau}\begin{pmatrix}
|\tau|^2&-\Re\tau\\
-\Re\tau&1
\end{pmatrix}
\;,
\qquad
\tau~=~ C_0+i e^{-\Phi}
\;,
\end{eqnarray}
the uplift formula~\eqref{DA} yields
\begin{eqnarray}
m_{\alpha\beta} &=&
\Delta^{4/3}\,{\cal S}_\alpha{}^a {\cal S}_\beta{}^b\,{\rm m}_{ab}
\;,
\label{dilax}
\end{eqnarray}
where $S$ is an $\textup{SO}(2)$ rotation matrix parameterised by
\begin{eqnarray}
{\cal S} &=&
\begin{pmatrix}
\cos\left(\frac34\,\varphi-\frac14\,\omega\right)
&
\sin\left(\frac34\,\varphi-\frac14\,\omega\right)
\\
-\sin\left(\frac34\,\varphi-\frac14\,\omega\right)
&
\cos\left(\frac34\,\varphi-\frac14\,\omega\right)
\end{pmatrix}
\;,
\label{SO2}
\end{eqnarray}
and ${\rm m}_{ab}$ is a ${GL}(2)$ matrix with entries
\begin{eqnarray}
{\rm m}_{11}&=&
\frac{1+\mu ^2 \nu ^2}{8\, \mu ^2 \nu ^4}
\Big(
\left(1+\nu ^4\right) \left(\mu ^2+\nu ^2\right)
+\left(1-\nu ^4\right) \left(\mu ^2-\nu ^2\right)\left(U^2-V^2\right)
\Big)
  \;,\nonumber\\
 {\rm m}_{12} &=&
\frac{ \left(1-\mu^2 \nu ^2\right) \left(1-\nu ^4\right) \left(\mu ^2+\nu ^2\right) }{4\, \mu ^2 \nu ^4}
  \left(U\cdot V\right)
\;,
 \nonumber\\
{\rm m}_{22} &=&
\frac{1+\mu ^2 \nu ^2}{8\, \mu ^2 \nu ^4}
\Big(
\left(1+\nu ^4\right) \left(\mu ^2+\nu ^2\right)
-\left(1-\nu ^4\right) \left(\mu ^2-\nu ^2\right)\left(U^2-V^2\right)
\Big)
\;.
\label{mm}
\end{eqnarray}

It is straightforward to check that the determinant of ${\rm m}_{ab}$ is given
by $\Delta^{-8/3}$~\eqref{warp} as required in order to have $m_{\alpha\beta}
\in \textup{SL}(2)$\,. Again, for vanishing angles we recover the result
from~\cite{Pilch:2000fu}.

It is remarkable that the dependence of the IIB axion/dilation on the (a priori
$x$-dependent) 5D angles $(\varphi,\omega)$ is entirely captured by a rotation
of the internal coordinates~\eqref{UVrotated} and the
$\textup{SO}(2)\subset\textup{SL}(2)_{\rm IIB}$
rotation~\eqref{SO2}\footnote{Note that when evaluated on~\eqref{lambda-source}
the $\textup{SO}(2)\subset\textup{SL}(2)_{\rm IIB}$ becomes the identity and
$\Theta^i =dU^i, \Lambda^i = dV^i$ (see~\eqref{ThetaLambda}) and thus the 10
dimensional solution involving a non-trivial $\lambda(x)$ can be obtained by
simply substituting $u^i, v^i$ with $U^i, V^i$.}. We will see in the following
that this feature persists for the full IIB uplift. In particular, $2\pi$
periodicity of the $5D$ theory implies that the IIB uplift is invariant under
the combination of an exchange $U^i \leftrightarrow V^i$ with a constant
$\textup{SL}(2)$ rotation~\eqref{SO2} with ${\cal S}=-i\sigma_2$, as is easily
verified for~\eqref{G5} and~\eqref{mm}.

\subsection{Uplift of the 4-scalar truncation: \texorpdfstring{$p$}{p}-forms}

We now evaluate~\eqref{C2C4} on the four scalar truncation in order to derive
the IIB $p$-forms. % For the two-form doublet, we find
\begin{eqnarray}
\label{fullpot}
C_{mn\,\alpha} &=& \Delta^{8/3}\,{\cal S}_{\alpha}{}^a\,{\rm C}_{mn\,a}
\;,
\end{eqnarray}
where ${\cal S}$ is the $\textup{SO}(2)$ rotation matrix in~\eqref{SO2}, and the 2-form
${\rm C}_a$ are
\begin{eqnarray}
{\rm C}_1 &=&
b_1\,\varepsilon^{ijk} \,
\Big(
\left(1-\mu ^2 \nu ^2\right)\left(\mu ^2+\nu ^6\right)
V^i\, \Theta^j \wedge\Theta^k
 +\left(1+\mu ^2 \nu^2\right) \left(\mu ^2-\nu ^6\right)
 V^i\, \Lambda^j \wedge\Lambda^k
\nonumber\\
&&{}\qquad\qquad
+ 2\,\nu^2\left(1-\mu ^4 \nu ^4\right) U^i \, \Theta^j \wedge \Lambda^k
 \Big)
 \nonumber\\
 &&{}
 + b_2\,\varepsilon^{ijk} \,
\Big(
\left(1+\mu ^2 \nu ^2\right)\left(\mu ^2-\nu ^6\right)
U^i\, \Theta^j \wedge\Theta^k
 +\left(1-\mu ^2 \nu^2\right) \left(\mu ^2+\nu ^6\right)
 U^i\, \Lambda^j \wedge\Lambda^k
\nonumber\\
&&{}\qquad\qquad
+ 2\,\nu^2 \left(1-\mu ^4 \nu ^4\right) V^i \, \Theta^j \wedge \Lambda^k
 \Big)
 \;,
\nonumber\\[1ex]
{\rm C}_2 &=&-{\rm C}_1\Big|_{U^i\leftrightarrow V^i, \Theta^i\leftrightarrow \Lambda^i}
\;,
\label{C2}
\end{eqnarray}
with the functions
\begin{eqnarray}
b_1 &=&
-\frac{1+\mu ^2 \nu^2}{64\, \mu ^4 \nu ^8}\,
\Big(\left(1+\nu ^4\right) \left(\mu ^2+\nu ^2\right)
+\left(1-\nu ^4\right)\left(\mu ^2-\nu ^2\right)\left(U^2-V^2\right)
\Big)  \;,
\nonumber\\
  b_2 &=&
  \frac{1-\mu ^2 \nu^2}{32\, \mu ^4 \nu ^8}\,\Big(
 \left(1-\nu ^4\right) \left(\mu ^2+\nu ^2\right)(U\cdot V)\Big)
\;.
\end{eqnarray}
Again, the dependence on the $5D$ angles $(\varphi,\omega)$ is entirely captured
by the $\textup{SO}(2)$ rotation~\eqref{SO2} and the rotated basis~\eqref{UVrotated}.
The internal component of the 4-form potential takes the form
\begin{eqnarray}
C &=& \mathring{C}+ \frac1{4!}\,\Delta^{8/3}\,\varepsilon_{ijm}\varepsilon_{kln}\left(f_1 \,(U^m U^n - V^m V^n) + 2\,f_2\,U^{(m}V^{n)}\right)
\Theta^i \wedge \Theta^j\wedge \Lambda^k\wedge \Lambda^l
\nonumber\\
&&{}
+\frac1{4!}\,\Delta^{8/3}\,f_3\,\Theta^i\wedge \Theta^j\wedge \Lambda^i\wedge \Lambda^j
\;,
\end{eqnarray}
where the background field $\mathring{C}$ is given in~\eqref{volume} and
\begin{eqnarray}
f_1 &=& -\frac{3\,(U\cdot V)}{32\, \mu ^4 \nu ^8}
\left(1-\nu ^4\right)^2 \left(\mu ^2+\nu ^2\right)^2 \left(1-\mu ^2
  \nu ^2\right)^2
  \;,\nonumber\\
   f_2 &=&
   \frac{3\left(U^2-V^2\right)}{64\, \mu ^4 \nu ^8}
   \left(1-\nu ^4\right)^2 \left(\mu ^2-\nu ^2\right)^2 \left(1+\mu ^2 \nu
  ^2\right)^2
  \;,\nonumber\\
f_3 &=&
\frac{3\,(U\cdot V) \left(U^2-V^2\right)}{8\,\mu ^2 \nu ^6}
\left(1-\nu ^4\right)^3 \left(1-\mu ^4\right)
  \;.
\end{eqnarray}
Finally, the external component $C_{\mu\nu\rho\sigma}$ is determined from
integrating the $y$-independent function\footnote{Here, and in the following, we
use the notation $\omega_{\mu\nu\rho\sigma\tau}=\sqrt{|
g|}\,\varepsilon_{\mu\nu\rho\sigma\tau}$ for the 5D volume form.}
\begin{eqnarray}
5\,\partial_{[\mu}\,C_{\nu\rho\sigma\tau]}
&=& -\frac1{3}\,\omega_{\mu\nu\rho\sigma\tau}\left(V_{\rm pot}
-\frac{1}{6}\,
 {F}_{\kappa\lambda}\,{F}^{\kappa\lambda}\right)
-\frac{20\,\sqrt{2}}{9}
 \, {F}_{[\mu\nu} F_{\rho\sigma}{} A_{\tau]}
\;,
\label{C4}
\end{eqnarray}
with the scalar potential $V_{\rm pot}$ from~\eqref{D5potential}.

\subsection{Five-form field strength and self-duality equations}

As a first consistency check, we compute the IIB 5-form field strength
\begin{eqnarray}
 {F}_{\hat{\mu}_1\ldots\hat{\mu}_5} &\equiv& 5\,\partial_{[\hat{\mu}_1}{C}_{\hat{\mu}_2\ldots \hat{\mu}_5]}
 -\frac{15}4 \,
 \varepsilon_{\alpha\beta}\,{C}_{[\hat{\mu}_1\hat{\mu}_2}{}^{\alpha}\partial_{\hat{\mu}_3} {C}_{\hat{\mu}_4\hat{\mu}_5]}{}^{\beta}\;,
 \label{F5}
\end{eqnarray}
and verify that it satisfies the first order self-duality equations
\begin{eqnarray}
 F &=& \star F
\;.
\label{self-dual}
\end{eqnarray}
Here and in the following, indices $\hat\mu$ refer to the ten-dimensional
coordinates, split as $\{x^{\hat\mu}\}=\{x^\mu, y^m\}$\,. After some computation
we find that the internal components ${F}_{m_1\dots m_5}$ calculated from the
above expressions for $C_{mn}{}^\alpha$ and $C_{klmn}$, take the compact form
\begin{eqnarray}
 {F}_{m_1\dots m_5} &=&
 -\frac13\,\Delta^{8/3}\,\mathring{\omega}_{m_1\dots m_5}\,V_{\rm pot}
 \;,
 \label{F55}
\end{eqnarray}
with $V_{\rm pot}$ the scalar potential in~\eqref{D5potential}, and
$\mathring{\omega}_{m_1\dots m_5}$ is the volume-form of $S^5$ defined in
equation~\eqref{eq:S5_volume_form}. The external component of the five-form
field strength $F_{\mu\nu\rho\sigma\tau}$ is computed
from~\eqref{C4},~\eqref{F5} as
\begin{eqnarray}
F_{\mu\nu\rho\sigma\tau} &=&
5\,\partial_{[\mu} C_{\nu\rho\sigma\tau]}
-10\,F_{[\mu\nu}{}^m C_{\rho\sigma\tau]m}
\nonumber\\
&=&
-\frac1{3}\,\omega_{\mu\nu\rho\sigma\tau}\left(V_{\rm pot}
-\frac{1}{12}\,
 {F}_{\kappa\lambda}\,{F}^{\kappa\lambda}\right)
-\frac{2}{9}
 \, {F}_{[\mu\nu} F_{\rho\sigma}{} A_{\tau]}
\nonumber\\
&&{}
+\frac{5}{16}\,\,F_{[\mu\nu}{}^{ga} {\cal Y}^g{\cal Y}^b
\left(
4\,\,\omega_{\rho\sigma\tau] \kappa\lambda}\,M_{ab,cd} F^{\kappa\lambda\,cd}
+3\,\sqrt{2}\,\varepsilon_{abcdef}\,
F_{\rho\sigma}{}^{cd} A_{\tau]}{}^{ef} \right)
\nonumber\\
&=&
-\frac1{3}\,\omega_{\mu\nu\rho\sigma\tau}\,V_{\rm pot}
\;,
\label{F5ext}
\end{eqnarray}
upon using~\eqref{vec_rel}. Comparing to~\eqref{F55}, we find that five-form we
have is indeed self-dual. Similar calculations lead to the other components of
the five-form
\begin{eqnarray}
  {F}_{\mu\, m_1\dots m_4} \,Dy^{m_1}\wedge Dy^{m_2}\wedge Dy^{m_3}\wedge Dy^{m_4}
  &=&
  3\,\Delta\,\star_5 \!\left(U^i\Lambda^i-V^i\Theta^i
\right)
\,{\cal J}_\mu
\;,
\label{F14}
\end{eqnarray}
where the current ${\cal J}_\mu$ is defined in~\eqref{5DMaxwell5} and the 5D
Hodge dual $\star_5$ gives explicitly
\begin{eqnarray}
2\,\Delta^{-5/3}\,\star_5\!\left(
U^i\Lambda^i-V^i\Theta^i
\right)
 &=&
\Theta^i \wedge\Theta^j \wedge\Lambda^i \wedge\Lambda^j
\nonumber\\
&&{}
+\frac{(1-\nu^4)^2}{\nu^4}\,\left(U^i U^j+V^i V^j\right)
\Theta^i \wedge\Theta^k\wedge \Lambda^j \wedge\Lambda^k
\;.
\end{eqnarray}
Comparing to the result for
\begin{eqnarray}
F_{\mu\nu\rho\sigma m}Dy^m &=&
\frac{1}{12} \,\omega_{\mu\nu\rho\sigma\lambda}\,
\left(
D_{\kappa} F^{\lambda\kappa}
+\frac{1}{6}\,\omega^{\lambda\nu_1\dots\nu_4}\,F_{\nu_1\nu_2}{} F_{\nu_3\nu_4}
\right)
\left(
U^i\Lambda^i-V^i\Theta^i
\right)
\;,
\end{eqnarray}
shows explicitly that the IIB self-duality equations~\eqref{self-dual} reduce to
the Maxwell equations of the $D=5$ theory (\ref{5DMaxwell5}). For the remaining
components, we obtain after some calculation
\begin{eqnarray}
F_{\mu\nu\rho mn}
&=&
-\frac{1}{12}\,\omega_{\mu\nu\rho\sigma\tau}\, F^{\sigma\tau}\,
 (\Theta^i\wedge \Lambda^i)_{mn}
\;,
\nonumber\\
F_{\mu\nu kmn}&=&
-\frac{\Delta^{1/3}}{12}\,F_{\mu\nu}\,\star_5\!\left(\Theta^i\wedge \Lambda^i\right)_{kmn}
\;,
\label{F53}
\end{eqnarray}
again in accordance with self-duality of the IIB field strength.

\subsection{Dual 6-forms}

For an explicit check of the remaining field equations, we further truncate down
to two real scalar fields, i.e.\ we assume constant angles and set the vector
field to zero, so that in particular the IIB metric is block diagonal. This is
precisely compatible with the GPPZ solution~\eqref{5rsol}. The ten-dimensional
IIB 3-form field equations take the form
\begin{eqnarray}
\nabla_{\hat\rho} \left( m^{\alpha\beta}
F^{\hat\mu\hat\nu\hat\rho}{}_{\beta} \right)
&=&
-\frac{2}{3}\,\varepsilon^{\alpha\beta} F^{\hat\mu\hat\nu\hat\kappa\hat\lambda\hat\rho} F_{\hat\kappa\hat\lambda\hat\rho}{\,}_\beta
\;,
\label{DF3}
\end{eqnarray}
and we have explicitly checked that they are verified if the five-dimensional
scalar fields satisfy the five-dimensional equations of motion induced
by~\eqref{D5_4scalars}. Rather than going through the details of this
calculation, let us give an equivalent consistency check by extracting the dual
six forms in ten dimensions. The field equations~\eqref{DF3} may be rewritten as
the Bianchi identities
\begin{eqnarray}
 \partial_{[\hat\rho_1} F_{\hat\rho_2 \dots \hat\rho_8]}{}^\alpha
&=&
28\,\varepsilon^{\alpha\beta}\,
F_{[\hat\rho_1 \dots \hat\rho_5} F_{\hat\rho_6\hat\rho_7\hat\rho_8]\,\beta}
\label{BianchiF7}
\end{eqnarray}
for the dual 7-form field strength $F_{\hat\rho_1 \dots \hat\rho_7}{}^\alpha$
defined by
\begin{eqnarray}
F_{\hat\rho_1 \dots \hat\rho_7}{}^\alpha &\equiv&
\frac16\,\sqrt{|G|}\,\varepsilon_{\hat\rho_1 \dots \hat\rho_7\hat\mu\hat\nu\hat\rho}\,m^{\alpha\beta}\,
F^{\hat\mu\hat\nu\hat\rho}{}_\beta\;.
\label{defF7}
\end{eqnarray}
The Bianchi identities~\eqref{BianchiF7} may then be integrated to
\begin{eqnarray}
F_{\hat\rho_1 \dots \hat\rho_7}{}^\alpha&=&
  7\,\partial_{[\hat\rho_1} C_{\hat\rho_2\dots\hat\rho_7]}{}^\alpha
 -84\,\varepsilon^{\alpha\beta}\,C_{[\hat\rho_1\hat\rho_2\,\beta} F_{\hat\rho_3 \dots \hat\rho_7]}
 \nonumber\\ &&{}
  -70\,\varepsilon^{\alpha\beta}\,\varepsilon^{\gamma\delta}\,C_{[\hat\rho_1\hat\rho_2\,\beta} C_{\hat\rho_3\hat\rho_4\,\gamma} F_{\hat\rho_5\hat\rho_6\hat\rho_7]\,\delta}
  \;,
\end{eqnarray}
in terms of the dual six-form gauge potential
$C_{\hat\rho_1\dots\hat\rho_6}{}^\alpha$\,. With the above explicit expressions
for the IIB gauge potentials~\eqref{C2} and field strengths
(\ref{F55}),~\eqref{F5ext} of our ten-dimensional solution, we find that
equations~\eqref{defF7} can explicitly be integrated to give the following
non-vanishing components of the six form
\begin{eqnarray}
C_{\mu\nu\rho\sigma\tau\,m}{}^\alpha&=&
\omega_{ \mu\nu\rho\sigma\tau}\,\Xi_{m}{}^\alpha
\nonumber\\
C_{\mu\nu\rho\sigma,\,mn}{}^\alpha&=& \omega_{ \mu\nu\rho\sigma\tau}\,g^{\tau\lambda}\,
\Xi_{\lambda\,mn}{}^\alpha\;,
\;,
\end{eqnarray}
in terms of the one- and two-forms
\begin{eqnarray}
\Xi^1 &=&-
\frac{(1-\mu^2\nu^2)\,\big((\mu^4-\nu^4)(1-\mu^2\nu^2)+2\,\mu^2\nu^2\,
(1+\mu^2\nu^2)\big)}{8\,\mu^4\,\nu^4}\,\varepsilon_{ijk}\,U^i\,V^j\,\Lambda^k
\;,\nonumber\\
\Xi^2 &=& \Xi^1\Big|_{U\leftrightarrow V,\Theta\leftrightarrow\Lambda}
\;,\nonumber\\[2ex]
\Xi_{\lambda}{}^{1}
&=&
\varepsilon_{ijk}\left(\frac{\partial_\lambda\nu}{\nu}-\frac{\partial_\lambda\mu}{3\mu} \right)
U^i \,\Theta^j\wedge\Theta^k
+\varepsilon_{ijk}\left(\frac{\partial_\lambda\mu}{\mu} +\frac{\partial_\lambda\nu}{\nu}\right)
U^i \,\Lambda^j\wedge\Lambda^k
\;,
\nonumber\\
\Xi_{\lambda}{}^{2} &=&
\Xi_{\lambda}{}^{1}\Big|_{U\leftrightarrow V,\Theta\leftrightarrow\Lambda}
\;.
\end{eqnarray}

\subsection{Einstein equations}

It remains to check the dilaton/axion equations and the Einstein equations. In
our IIB conventions, these read
\begin{eqnarray}
\nabla_{\hat{\mu}} \left(  m^{\beta\gamma} \partial^{\hat{\mu}}m_{\alpha\gamma} \right)
&=&
-\frac{1}{6} \,m^{\beta\gamma} {F}_{\hat{\mu}_1\hat{\mu}_2\hat{\mu}_3\,\gamma}{F}^{\hat{\mu}_1\hat{\mu}_2\hat{\mu}_3}{}_{\alpha}
+\frac{1}{12}\, \delta_\alpha{}^\beta\,m^{\gamma\delta} {F}_{\hat{\mu}_1\hat{\mu}_2\hat{\mu}_3\,\gamma}{F}^{\hat{\mu}_1\hat{\mu}_2\hat{\mu}_3}{}_{\delta}
\;,
\end{eqnarray}
and
\begin{eqnarray}
R_{\hat\mu\hat\nu}-\frac12\,G_{\hat\mu\hat\nu}\,R &=&
\frac16\,F_{\hat\mu\hat\kappa\hat\lambda\hat\sigma\hat\tau}{} F_{\hat\nu}{}^{\hat\kappa\hat\lambda\hat\sigma\hat\tau}
+
\frac14\,F_{\hat\mu\hat\sigma\hat\tau}{}^\alpha F_{\hat\nu}{}^{\hat\sigma\hat\tau}{}^\beta m_{\alpha\beta}
-\frac1{24}\,G_{\hat\mu\hat\nu}\,F_{\hat\rho\hat\sigma\hat\tau}{}^\alpha F^{\hat\rho\hat\sigma\hat\tau}{}^\beta m_{\alpha\beta}
\nonumber\\
&&{}
 -\frac14\,\partial_{\hat\mu} m_{\alpha\beta} \partial_{\hat\nu} m^{\alpha\beta}
+\frac18\,G_{\hat\mu\hat\nu}\,\partial_{\hat\rho} m_{\alpha\beta} \partial^{\hat\rho} m^{\alpha\beta}
\;.
\label{Einstein}
\end{eqnarray}
It is a tedious computation to check that with the above expressions for the
metric~\eqref{metricIIBexplicit}, dilaton-axion matrix~\eqref{dilax}, gauge
potentials~\eqref{C2} and field strengths (\ref{F55}),~\eqref{F5ext}, these
field equations are indeed satisfied. We have explicitly verified all components
of these equations using Mathematica~\cite{Mathematica}. We attach a Mathematica
file with the explicit IIB solution and the verification of Einstein's equations
in the arXiv version of this paper. Let us just note that the contribution from
the five-form field strength to the energy-momentum tensor on the r.h.s.\
of~\eqref{Einstein} is simply given by
\begin{eqnarray}
 F_{\mu}{}^{\rho_1\rho_2\rho_3\rho_4}\,F_{\nu \rho_1\rho_2\rho_3\rho_4}
&=&
-\frac{8}3\, \Delta^{10/3}\,V_{\rm pot}^2\,G_{\mu\nu}
\;,\nonumber\\
F_{m}{}^{k_1k_2k_3k_4}\,F_{nk_1k_2k_3k_4}
&=&\frac{8}3\,\Delta^{10/3}\,V_{\rm pot}^2\,G_{mn}
\;.
\end{eqnarray}
In contrast, the remaining terms on the r.h.s.\ of~\eqref{Einstein} produce very
lengthy expressions in the scalars $\mu$, $\nu$, their derivatives, and the
internal coordinates $U^i$, $V^i$, which we do not report in detail. They
combine however precisely into the Einstein tensor computed from the
metric~\eqref{metricIIBexplicit} upon using the first order flow
equations~\eqref{flowmunu}. All the ten-dimensional equations are thus
satisfied.

%%%%%%%%%%%%%%%%%%%%%%%%%%%%%%%%%%%%%%%%%%%%%%%%%%%%
%
% UV asymptotics of the uplifted solution
%
%%%%%%%%%%%%%%%%%%%%%%%%%%%%%%%%%%%%%%%%%%%%%%%%%%%%
\section{UV asymptotics of the uplifted solution}%
\label{sec:asymptotics}

In order to interpret our ten-dimensional solution we can compute its asymptotic
behaviour for large values of the radial coordinate and check whether the
various field have the fall-off expected from the AdS/CFT dictionary. It is also
interesting to compare our results with the asymptotic behaviours of the other
supergravity solutions that are supposed to describe $\mathcal{N}=1^*$, namely
the Polchinski-Strassler solution~\cite{Polchinski:2000uf} and the
zero-temperature limit of the Freedman-Minahan solution~\cite{Freedman:2000xb}.
To this purpose we perform the change of variable $t = e^{C_1}/r$, where $r$ is
the radial coordinate used in~\cite{Polchinski:2000uf,Freedman:2000xb}. In this
section we only give terms up to quadratic order in the deformation parameters
$m$ and $\sigma$ and we fix the values of the angles $\varphi$ and $ \omega$ to
constant values.
%Other constant values of
$\varphi$ and $ \omega$ correspond to rotations in the space of harmonics. More
general expressions can be found in Appendix~\ref{app:asymptotics}.

\paragraph{Dilaton/axion}
The expansion takes a particularly simple form for the
field $B = \frac{1 + i \tau}{1 - i \tau}$ that appears in the Kaluza-Klein
expansion around $S^5$~\cite{Kim:1985ez}. The first terms can be easily
computed for any value of the angles $\varphi$ and $ \omega$ and are
\begin{eqnarray}
\label{axiodilUV}
B &\sim & -  \left( \frac{ 4 m_0^2}{9 r^2 }e^{-2 i \varphi}  -\frac{ 2 m_0 \sigma_0}{3 \sqrt{3} r^4} e^{- i (\varphi - \omega)} \right)
(u^2 - v^2) \nonumber \\
&& - 2 i \left( \frac{ 4 m_0^2}{9 r^2 }e^{-2 i \varphi}  + \frac{ 2 m_0 \sigma_0}{3 \sqrt{3} r^4}  e^{- i (\varphi - \omega)} \right) (u \cdot v) \, .
\end{eqnarray}
The constants $m_0$ and $\sigma_0$ are given in~\eqref{eq:gppz_asymptotics} and
are related to the UV mass deformation and the expectation value of gaugino
condensate. The functions $(u^2 - v^2)$ and $(u \cdot v)$ corresponds to
$\textup{SO}(3)$ invariant scalar harmonics $Y^{(2,0)}$ and $Y^{(2,1)}$ on $S^5$
in the ${\bf 20}$ of $\textup{SO}(6)$ (see Appendix~\ref{app:harmonics}).

From~\eqref{axiodilUV}, setting the angles $\varphi$ and $ \omega$ to zero, we
can compute the expansions of the dilaton and axion
\begin{eqnarray}
\label{UVdil}
e^\Phi &\sim & 1 + \frac{2}{3} \frac{ m_0^2}{r^2}  (u^2 - v^2) +  \frac{m_0 \sigma_0 }{\sqrt{3}r^4}  (u^2 - v^2)  \, ,   \\
C_0 &\sim &  \frac{4}{3} \frac{m_0^2}{r^2}  (u \cdot v) + \frac{2}{\sqrt{3}} \frac{m_0 \sigma_0 }{r^4}   (u \cdot v)  \, .
\end{eqnarray}
The leading behaviour of the dilaton is the same as for the zero-temperature
limit of the Freedman-Minahan solution~\cite{Freedman:2000xb}. As already
discussed in~\cite{Freedman:2000xb}, its behaviour does not agree with the
asymptotic limits of the dilaton in the Polchinski-Strassler
solution~\cite{Polchinski:2000uf}, where the leading term of the dilaton is in a
singlet of $\textup{SO}(6)$.

\paragraph{Metric}
For $\varphi = \omega =0$ the large $r$ behaviour of the
metric is
\begin{eqnarray}
\label{UVmet}
d s^2_{10} = r^2 \left( 1+ \frac{m_0^2}{24 r^2} \right) d s^2_4 + \frac{d r^2}{r^2} \left( 1+ \frac{m_0^2}{16 r^2} \right) + d s^2_5
\end{eqnarray}
where $d s^2_4$ is the flat Minkowski metric in four dimensions and the internal
metric $d s^2_5$ is given by
\begin{eqnarray}
\label{UVintmet}
d s^2_5 &=&  (d u_i)^2 + ( d v_i)^2 + d u_i d u_j \left( \frac{m_0^2}{r^2} A_{ij} +  \frac{m_0 \sigma_0}{r^4} B_{ij} \right) \nonumber \\
&& + d v_i d v_j \left( \frac{m_0^2}{r^2} C_{ij} +  \frac{m_0 \sigma_0}{r^4} D_{ij} \right)
+ d u_i d v_j \left( \frac{m_0^2}{r^2} E_{ij} + \frac{m_0 \sigma_0}{r^4} F_{ij}  \right)
\end{eqnarray}
with coefficients
\begin{eqnarray}
A_{ij} &=& - \frac{1}{6} ( 3 + 4 (u^2 - v^2))  \delta_{ij} +  \frac{1}{3}  v_i v_j \, , \nonumber \\
B_{ij} &=& \frac{1}{\sqrt{3}} [ (u^2 -v^2)  \delta_{ij} +  v_i v_j  ] \, , \\
C_{ij} &=&  - \frac{1}{6}  ( 3 - 4 (u^2 - v^2))  \delta_{ij}  + \frac{1}{3}   u_i u_j  \, , \\
D_{ij} &=&  \frac{1}{\sqrt{3}}  [ - (u^2 -v^2) ] \delta_{ij}  +     u_i u_j ] \, , \\
E_{ij}&=& \frac{1}{3} [ -8 (u \cdot v) \delta_{ij} - 8 u_i v_j + 6 v_i u_j] \, , \\
F_{ij} &=&  \frac{2}{\sqrt{3}} [ - 2 (u \cdot v) \delta_{ij}  + u_i v_j + v_i u_j ] \, .
\end{eqnarray}

The form of the metric agrees with the structure of the zero-temperature limit
of the Freedman-Minahan solution.

\paragraph{Two-form potentials and field strengths}
For $\varphi=\omega =0$, the
first terms in the expansion of the two-forms potentials~\eqref{fullpot} are
\begin{eqnarray}
\label{UVas2pot}
C_1 &=& \frac{1}{2} \left( \frac{1}{\sqrt{3} } \frac{m_0}{r} + \frac{1}{2} \frac{\sigma_0}{r^3} \right)  \epsilon_{ijk}  v_i d u_j \wedge d u_k + \frac{1}{2} \left( \sqrt{3} \frac{m_0}{r} - \frac{1}{2} \frac{\sigma_0}{r^3} \right)  \epsilon_{ijk}  v_i d v_j \wedge d v_k \, , \nonumber \\
&& +  \left( \frac{1}{\sqrt{3} } \frac{m_0}{r} + \frac{1}{2} \frac{\sigma_0}{r^3} \right)  \epsilon_{ijk}  u_i d u_j \wedge d v_k  \\
C_2 &=&  -  \frac{1}{2} \left( \sqrt{3} \frac{m_0}{r} - \frac{1}{2} \frac{\sigma_0}{r^3} \right)  \epsilon_{ijk}  u_i d u_j \wedge d u_k
- \frac{1}{2} \left( \frac{1}{\sqrt{3} } \frac{m_0}{r} + \frac{1}{2} \frac{\sigma_0}{r^3} \right)  \epsilon_{ijk}  u_i d v_j \wedge d v_k \nonumber \\
&& -  \left( \frac{1}{\sqrt{3} } \frac{m_0}{r} + \frac{1}{2} \frac{\sigma_0}{r^3} \right)  \epsilon_{ijk}  v_i d u_j \wedge d v_k  \, .
\end{eqnarray}
A simple derivation gives the asymptotic behaviour of the fully internal
components of the field strengths $F_i = d C_i$
\begin{eqnarray}
\label{3ffluxas}
F_1 &=&  \frac{3}{2}  \left( \frac{1}{\sqrt{3} } \frac{m_0}{r}  + \frac{1}{2} \frac{\sigma_0}{r^3} \right)  \epsilon_{ijk}  d u_i \wedge d u_j \wedge d v_k  +
 \left( 3 \sqrt{3} \frac{m_0}{r}  - \frac{3}{2} \frac{\sigma_0}{r^3} \right)  d v_1  \wedge d v_2 \wedge d v_3  \\
F_2 &=&  - \frac{3}{2}  \left ( \frac{1}{\sqrt{3} } \frac{m_0}{r} + \frac{1}{2} \frac{\sigma_0}{r^3} \right) \epsilon_{ijk}  d u_i \wedge d v_j \wedge d v_k -
\left ( 3 \sqrt{3} \frac{m_0}{r} - \frac{3}{2} \frac{\sigma_0}{r^3} \right)  d u_1  \wedge d u_2 \wedge d u_3  \nonumber
\end{eqnarray}
The terms in $1/r$ in the~\eqref{UVas2pot} reproduce the large $r$ behaviour of
the two-form potentials of the Polchinski-Strassler
background~\cite{Polchinski:2000uf}, but the leading terms in $\sigma_0$
disagree. For a different values of the angles, $\varphi = \pi/2$ and a constant
arbitrary $\omega$ we also recover the leading behaviour $T=0$ limit of the
three-forms in the~\cite{Freedman:2000xb}.

\paragraph{Five-form flux}
Using~\eqref{F55} it is easy to derive the large $r$
behaviour of the purely internal component of five-form flux \begin{eqnarray}
F_5 = - \frac{1}{5!} \left( \frac{4}{r^6} - \frac{ 12 m_0^2}{r^8} \right)
\epsilon_{m_1 \ldots m_6} y^{m_1} dy^{m_2} \wedge dy^{m_3} \wedge  dy^{m_4}
\wedge dy^{m_5} \wedge dy^{m_6} \end{eqnarray} where $y^m$ are the six
coordinates of $R^6$ that parameterise the internal manifold. Again this
expression agrees with what given by Freedman and Minahan.

%%%%%%%%%%%%%%%%%%%%%%%%%%%%%%%%%%%%%%%%%%%%%%%%%%%%
%
% Singularity
%
%%%%%%%%%%%%%%%%%%%%%%%%%%%%%%%%%%%%%%%%%%%%%%%%%%%%
\section{Singularity}%
\label{sec:singularity}
In this section we discuss the behaviour of the ten-dimensional solution as we
approach the position where the five-dimensional solution has a curvature
singularity.

As we reviewed in \textsection~\ref{sec:GPPZ}, in five dimensions we have a
metric coupled to two real scalars $m$ and $\sigma$, which have a domain-wall
profile along the radial direction. The complete solution is given
in~\eqref{5rsol} and contains two integration constants $C_1$ and $C_2$, which
parametrize the mass deformation and the gaugino condensate, respectively. The
geometry is singular as the radial coordinate $y$ approaches either $C_1$ or
$C_2$, as one can verify by computing curvature invariants. As
in~\cite{Pilch:2000fu} we parametrize the location of the singularity by
defining
\begin{equation}
t = \exp \left(-(y-C_1)\right), \qquad \chi = 2 \sqrt{1-t} \, .
\end{equation}
The singularity of the $5d$ metric is then located at $ t \to 1$ or $\chi \to
0$, and the curvature scalar and Kretschmann invariant are given by
\begin{equation}
R = -\frac{16}{\chi^4} + O(\chi^{-2}), \qquad R_{\mu\nu\rho\sigma} R^{\mu\nu\rho\sigma} = \frac{640}{\chi^8} + O(\chi^{-6}) \, .
\end{equation}
Since the scalar fields are also singular in this limit one may wonder if there
is a different conformal frame than the Einstein frame, where the geometry is
regular or at least less singular. It turns out that this is not the case. We
will see later that the situation is different in 10 dimensions. We also define,
again following~\cite{Pilch:2000fu},
\begin{equation}
\lambda = e^{-3(C_1 - C_2)} \, .
\end{equation}
It was argued in~\cite{Girardello:1999bd} that $C_1 \geq C_2$ and this
translates to $\lambda \leq 1$ with the equality corresponding to the case where
the singularities in $m(y)$ and $\sigma(y)$ coincide. The singularity structure
of the $10d$ solution depends of whether $\lambda <1$ or $\lambda=1$ and we will
discuss the two cases separately.

\subsection{The \texorpdfstring{$\lambda <1$}{lambda < 1} Case}

We have computed the curvature scalar of the $10d$ solution and this has a
regular limit as $\chi \to 0$, %provided $\lambda <1$,
\begin{equation}%
	\label{eq:ricci_scalar_full}
\lim_{\chi\to0}R^\text{(full)} = \frac{\sqrt{2}}{1-\lambda^2}\frac{3(1+\lambda^2) - w_1^2(1-4\lambda+\lambda^2) - w_2^2(1+4\lambda+\lambda^2)}{(1-w_1^2-w_2^2)^{5/4}}.
\end{equation}
However, the Ricci scalar is now singular at
\begin{equation}
\zeta \equiv 1-w_1^2-w_2^2=0
\end{equation}
which is precisely the ring singularity discussed in~\cite{Pilch:2000fu}.

The metric near the singularity is given by
\begin{equation}
\begin{split}
ds^2 &= \frac{\zeta^{1/4}}{\sqrt2}\Biggl\{
(1+\chi^2 f_0)(2e^{2C_1}(1-\lambda^2)^{1/3}\eta_{\mu\nu}dx^\mu dx^\nu + d\chi^2) + \frac{\chi^2}{4}\left(\frac{1}{\zeta}{Y_v^{(0,0)}}^2+\sigma_2^2 + \sigma_3^2\right) \label{eq:pw_metric} \\
&\quad{} + \frac{1}{2\zeta}\left(\frac{1-\lambda}{1+\lambda}dw_1^2 + \frac{1+\lambda}{1-\lambda}dw_2^2\right) + \frac{\chi^2}{8\zeta(1-\lambda^2)}\omega_\parallel + \frac{\chi^2}{16\zeta^2(1-\lambda^2)^2}\omega_D
\Biggr\} + \mathcal O(\chi^4)
\end{split}
\end{equation}
where $Y_v^{(0,0)}$ is an $\textup{SO}(3)$ vector harmonic, (see
Appendix~\ref{app:harmonics}). The coefficient $f_0$ and the differentials
$\omega_\parallel$ and $\omega_D$ are given by
\begin{align}
f_0 &= \lambda\frac{w_1^2 - w_2^2}{4\zeta(1-\lambda^2)} \\
\omega_\parallel &= (1+\lambda^2)(2\zeta+1)2e^{2C_1}(1-\lambda^2)^{1/3}\eta_{\mu\nu}dx^\mu dx^\nu + (2+(3\zeta-1)(1-\lambda^2))d\chi^2 \\
\begin{split}
\omega_D &=
\left[(1-\lambda)^2d(w_1^2) + (1+\lambda)^2d(w_2^2)\right]^2 \\
&\quad{} - \left[(1+\lambda^2)(3-2\zeta)-2\lambda(w_1^2 - w_2^2)\right]\left[(1-\lambda)^2dw_1^2 + (1+\lambda)^2 dw_2^2\right] \\
&\quad{} + 8\lambda(\zeta + 1)\left[(1-\lambda)^2 dw_1^2 - (1+\lambda)^2 dw_2^2\right] \, .
\end{split}
\end{align}

Pilch and Warner also computed the near-singularity metric
in~\cite{Pilch:2000fu}. Their metric is reproduced by setting $f_0 =
\omega_\parallel = \omega_D = 0$ and $\frac{1}{\zeta}Y_v^{(0,0)} \to \sigma^1$
(modulo a typo in one of the coefficients of $dw_1^2$). Note that $d\chi^2 +
\frac14\chi^2(\sigma_1^2 + \sigma_2^2 + \sigma_3^2)$ is just the flat metric on
$R^4$ and the terms in the first line of~\eqref{eq:pw_metric} combine to give
eight dimensional Minkowski spacetime. This was interpreted
in~\cite{Pilch:2000fu} as evidence that the singularity is associated with
7-branes. We cannot however ignore the terms with $f_0, \omega_\parallel,
\omega_D$ and $\frac{1}{\zeta}Y_v^{(0,0)}$ because they are of the same order as
$\frac14\chi^2(\sigma_1^2 + \sigma_2^2 + \sigma_3^2)$. Taking these terms into
account, we find no evidence for 7-branes in the near-singularity structure of
the metric. At the position of the $5d$ singularity, $\chi=0$, the 10$d$ metric
is of co-dimension 4:
\begin{equation}
ds^2 = \frac{\zeta^{1/4}}{\sqrt{2}} \left(2e^{2C_1}(1-\lambda^2)^{1/3}\eta_{\mu\nu}dx^\mu dx^\nu + \frac{1}{2\zeta}\left(\frac{1-\lambda}{1+\lambda}dw_1^2 + \frac{1+\lambda}{1-\lambda}dw_2^2\right) \right).
\end{equation}

Note that the limit $\chi \to 0$ is not a decoupling limit, i.e.\ the metric
in~\eqref{eq:pw_metric} does not solve the bulk equations of motion and its
curvature does not agree with (\ref{eq:ricci_scalar_full}). To properly account
for~\eqref{eq:ricci_scalar_full} one needs to keep higher order terms in $\chi$.
First consider the ten-dimensional metric $G_{MN}$. One can check that in the
expansion in the radial coordinate $\chi$ around $\chi=0$ the lowest order in
$\chi$ that occurs is the constant order $\chi^0$, which is also manifest
in~\eqref{eq:pw_metric}. The same analysis performed on the full inverse metric
$G^{MN}$ shows that its lowest order in $\chi$ is the order $\chi^{-2}$. Given
this information we can deduce to which orders we need to expand $G_{MN}$ and
$G^{MN}$ in order to obtain results consistent with the full computation of the
Ricci scalar in the limit $\chi\to0$. Schematically the Riemann tensor and the
Ricci scalar are given in terms of the metric and its inverse as follows:
\begin{align}
& R_{MNRS} \sim \partial^2 G + G^{-1}\partial G \partial G \label{eq:riemann_schematic} \\
& R \sim G^{-1}G^{-1}(\partial^2 G + G^{-1}\partial G \partial G).
\end{align}
Since $G^{-1}\sim 1/\chi^2$ to get the correct constant term in the Ricci scalar
the Riemann tensor needs to be at least of order $\chi^4$. Then, from the second
term in~\eqref{eq:riemann_schematic} we infer that $\partial G \partial G$ has
to be at least of order $\chi^6$, and since a derivative with respect to $\chi$
lowers the order in $\chi$ by 1, $G$ has to be at least of order $\chi^7$.
Similarly, one can deduce that one needs to keep terms at least up to order
$\chi^4$ in the inverse metric $G^{-1}$. We have explicitly checked that keeping
the metric and inverse metric to these orders one indeed obtains a curvature
scalar consistent with~\eqref{eq:ricci_scalar_full}.

Similarly, one can study the order to which one has to keep the other fields in
order for the bulk equations to be satisfied, order by order in $\chi^2$. In
general, one cannot truncate this series at some fixed order and have the field
equations satisfied, as different orders contribute to different terms in the
field equations.

We now provide the near-singularity behaviour of the warp factor and all other
fields.
\paragraph{Warp-Factor}
Following~\cite{Pilch:2000fu} we define
\begin{equation}
\xi^2 = \Delta^{-8/3} \, .
\end{equation}
Then the warp factor has the following leading behaviour as $\chi \to 0$,
\begin{equation}
\xi %= \frac{\zeta^{1/2}}{2(t-1)^2} + \mathcal O(t-1)
= \frac{8\zeta^{1/2}}{\chi^4} + \mathcal O(\chi^3) \, .
\end{equation}
\paragraph{Axion/Dilaton}
In the limit $\chi \to 0$ the axion/dilaton matrix $m_{\alpha\beta}$ is regular
and takes the following form:
\begin{equation}
m_{\alpha\beta} = \zeta^{-1/2}
\begin{pmatrix}
1+w_1 & w_2 \\
w_2 & 1-w_1
\end{pmatrix} + \mathcal O(\chi^2) \, .
\end{equation}

\paragraph{2-Form Potential}
In the limit $\chi \to 0$ the 2-form reduces to the following expression:
%One may write the limits in terms of wedge products of some vector harmonics found in Appendix~\ref{app:harmonics}:
\begin{align}
C^1 &=
\frac{\sqrt3}{\zeta}\left(Y_v^{(0,0)}\wedge Y_v^{(1,1)} - \frac12 dw_1\wedge Y_v^{(1,0)} - \frac12 dw_2\wedge Y_v^{(1,1)}\right)
+ \mathcal O(\chi^2) \\
C^2 &=
\frac{\sqrt3}{\zeta}\left(-Y_v^{(0,0)}\wedge Y_v^{(1,1)} + \frac12 dw_1\wedge Y_v^{(1,1)} - \frac12 dw_2\wedge Y_v^{(1,0)}\right)
+ \mathcal O(\chi^2)
\end{align}
where the $\textup{SO}(3)_\text{diag}$ invariant vector harmonics $Y_v^{(k,m)} = Y_{v\,
n}^{(k,m)}dy^n$ are 1-forms on the $S^5$ cotangent bundle, and $k$ and $m$ are
integers that label the harmonics (see Appendix~\ref{app:harmonics} for more
details).

\paragraph{4-Form Potential}
The limit $\chi \to 0$ for the 4-form is regular and gives the following result
\begin{equation}
C = \mathring{C} + \frac1{32} \frac{w_2}{\sqrt\zeta (1+w_1)} \; dw_1\wedge dw_2\wedge \sigma^2\wedge \sigma^3 + \frac{1}{16}(w_1 dw_2 - w_2 dw_1)\wedge \sigma^1 \wedge \sigma^2\wedge \sigma^3 + \mathcal O(\chi^2)
\end{equation}

\paragraph{6-Form Potential}
We get
\begin{equation}
C_{(6)} \to \frac{\sqrt3}{2} e^{4 C_1}(1-\lambda^2)^{2/3} d\tau \wedge dx^1 \wedge dx^2 \wedge dx^3 \wedge (\chi d\chi) \wedge
\begin{pmatrix}
Y_v^{(1,1)} \\ -Y_v^{(1,0)}
\end{pmatrix} + \mathcal O(\chi^2) \, .
\end{equation}
The coordinate $\tau$ denotes the time. Note that there are powers of $\chi$
that come from the volume form $\omega_{\mu\nu\rho\sigma\tau}$ and cancel some
of the divergences. See equation~\eqref{eq:AdS_volume_form} for the expansion of
$\omega_{\mu\nu\rho\sigma\tau}$ in terms of the radial coordinate.

\subsection{The \texorpdfstring{$\lambda = 1$}{lambda = 1} Case}
We now set $\lambda = 1$ first and then take the $\chi \to 0$ limit. The Ricci
scalar becomes
\begin{equation}
R^{(\lambda = 1)} = \frac{1}{6\sqrt{3}} \left(\frac{8}{\chi^2} - 1\right) \frac{(8 + w_1^2 - 8 w_2^2) (10 - w_1^2 - 10w_2^2)}{(4 - w_1^2 - 4w_2^2)^{9/4}} + \mathcal O(\chi).
\end{equation}
Thus in this case the $10d$ metric is still singular at $\chi=0$, though
diverging at slower rate than the $5d$ solution. In addition, the metric is
singular at $(w_1, w_2) = (0, \pm 1)$ (which corresponds to $(\theta, \phi) =
(\pi/4, \pi/2 \pm \pi/2)$), but there is no ring singularity anymore. The metric
itself takes the following form
\begin{equation}
\begin{split}
ds_5^2 &= \Omega^{1/2}\left[12^{1/3} e^{2 C_1}\eta_{\mu\nu}dx^\mu dx^\nu \chi^{2/3} + \left(\frac{3}{8}\chi^2+1\right)d\chi^2 + \frac{1}{24 \Omega ^2}\left(\frac{8}{\chi^2}-1\right)dw_2^2 +\frac{\chi^2}{4 \zeta}\delta_2 \right] \\
&\qquad{}+\mathcal O(\chi^{7/3}) \, .
\end{split}
\end{equation}
The differential $\delta_2$ is given by
\begin{equation}
\begin{split}
\delta_2 &= {Y_v^{(0,0)}}^2
+\frac{3}{4}dw_1^2
-\frac{1}{4}dw_2^2
-\frac{\zeta}{48 \Omega^2}dw_2^2
-\frac{\zeta(1+w_2^2)}{18 \Omega^4}dw_2^2
+ \frac{2-w_1^2}{6 \Omega^2}dw_2^2
+\frac{w_1 w_2}{2 \Omega^2}dw_1 dw_2 \\
&\quad{}
+\frac{w_1}{3 \Omega^2}Y_v^{(0,0)} dw_2
+\frac{9}{2+w_1} {Y_v^{(1,0)}}^2
+\frac{2+w_1}{\Omega^2}\left(\frac{2w_2}{2+w_1}Y_v^{(1,0)} - Y_v^{(1,1)}\right)^2 \, .
\end{split}
\end{equation}
The leading order terms in this metric reproduce the result found by Pilch and
Warner~\cite{Pilch:2000fu}, but we have additional subleading terms.

\paragraph{Warp-Factor}
The warp factor has the following leading behaviour under $\lambda\to 1$, then $\chi \to 0$
%(see the end of Section~\ref{sec:definitions} for the definition of $\hat\Omega$)
\begin{equation}
\xi %= \frac{\hat{\Omega}}{(t-1)^2} + \mathcal O(t-1)
= \frac{16\hat{\Omega}}{\chi^4} + \mathcal O(\chi^{-3}) \, ,
\end{equation}
where
\begin{equation}
\hat \Omega =\frac13\sqrt{4-w_1^2-4w_2^2} \, .
\end{equation}
\paragraph{Axion/Dilaton}
The axion/dilaton matrix $m_{\alpha\beta}$ is regular and takes the following
form:
\begin{equation}
m_{\alpha\beta} = \frac{1}{3\hat\Omega}
\begin{pmatrix}
2+w_1 & 2w_2 \\
2w_2 & 2-w_1
\end{pmatrix} + \mathcal O(\chi^2).
\end{equation}

\paragraph{2-Form Potential}
The limit of the 2-form may be written in terms of wedge products of the vector
harmonics found in Appendix~\ref{app:harmonics}:
\begin{align}
C^1 &=
\frac{\sqrt3}{\zeta}\left(Y_v^{(0,0)} - \frac{2\zeta - 3w_1}{18\hat\Omega^2}dw_2\right)\wedge Y_v^{(1,1)}
- \frac{\sqrt3}{2\zeta}\left(dw_1 + \frac{w_1 w_2}{3\hat\Omega^2}\right)\wedge Y_v^{(1,0)}
+ \mathcal O(\chi^2) \, , \\
C^2 &=
\frac{\sqrt3}{\zeta}\left(-Y_v^{(0,0)} - \frac{2\zeta + 3w_1}{18\hat\Omega^2}dw_2\right)\wedge Y_v^{(1,0)}
+ \frac{\sqrt3}{2\zeta}\left(dw_1 + \frac{w_1 w_2}{3\hat\Omega^2}\right)\wedge Y_v^{(1,1)}
+ \mathcal O(\chi^2)
\end{align}

\paragraph{4-Form Potential}
The limit of the 4-form is regular, and is given by
\begin{align}
\begin{split}
C &= \mathring{C} + \frac{w_2 \zeta^{1/2}}{72(1+w_1)\hat\Omega^2} \; dw_1\wedge dw_2\wedge \sigma^2\wedge \sigma^3 - \frac{w_1}{48\hat\Omega^2}dw_2\wedge \sigma^1\wedge \sigma^2\wedge \sigma^3 \\
&\qquad{} + \frac{1}{16}(w_1 dw_2 - w_2 dw_1)\wedge \sigma^1 \wedge \sigma^2\wedge \sigma^3 + \mathcal O(\chi^2)
\end{split} \\
\begin{split}
&= \mathring{C} + \left(\frac{1+w_1}{2}\right)^{-1}\frac{w_2 \zeta^{1/2}}{144\hat\Omega^2} \; dw_1\wedge dw_2\wedge \sigma^2\wedge \sigma^3 - \frac{w_1}{48\hat\Omega^2}dw_2\wedge \sigma^1\wedge \sigma^2\wedge \sigma^3 \\
&\qquad{} + \frac{1}{16}(w_1 dw_2 - w_2 dw_1)\wedge \sigma^1 \wedge \sigma^2\wedge \sigma^3 + \mathcal O(\chi^2) \, .
\end{split}
\end{align}

\paragraph{6-Form Potential}
For the 6-form we get
\begin{align}
C_{(6)} &\to -\frac{7 e^{4 C_1}}{2^{2/3} 3^{5/6}} d\tau\wedge dx^1 \wedge dx^2 \wedge dx^3 \wedge (\chi^{7/3}d\chi) \wedge
\begin{pmatrix}
Y_v^{(1,1)} \\ -Y_v^{(1,0)}
\end{pmatrix} + \mathcal O(\chi)^{10/3},
\end{align}
where $\tau$ is the time coordinate. Again, as in the case $\lambda < 1$, there
are some powers of $\chi$ that come from the volume form
$\omega_{\mu\nu\rho\sigma\tau}$ and cancel some divergences. See
equation~\eqref{eq:AdS_volume_form} for the expansion of
$\omega_{\mu\nu\rho\sigma\tau}$ in terms of the radial coordinate.

\subsection{Different frames}

Since the solution involves non-trivial scalars there is an intrinsic ambiguity
in the definition of the spacetime metric: one can rescale the metric with
powers of the scalars. Different probe behaviours see different metrics and
different conformal frames carry different physical meaning. For example,
supergravity probes see the Einstein frame metric and strings see the string
frame metric. In some cases singular geometries are regular in a different
frame. For example, the geometry of non-conformal D$p$ branes is singular in the
Einstein and string frame but it is regular in the ``dual
frame''~\cite{Duff:1994fg} and this is also the frame best suited for
holography~\cite{Boonstra:1998mp, Kanitscheider:2008kd}. Here we want to analyse
the dependence of the singularity on the choice of frame.

Usually one uses the dilaton when discussing different frames.\footnote{One
reason for this is that the axion is more properly viewed as a 0-form potential
and has an associated gauge invariance.} Since our solution has both an axion
and a dilaton we will explore a general rescaling by both: $g_{\hat\mu\hat\nu}
\to \tilde g_{\hat\mu\hat\nu} = \Omega^2 g_{\hat\mu\hat\nu}$ with the scaling
factor $\Omega = e^{x \phi} C_0^z$ given by some powers of the dilaton
$e^{\phi}$ and the axion $C_0$ parameterised through constants $x$ and $z$.
Given the definition of the axion/dilaton matrix $m_{\alpha\beta}$
\begin{equation}
m_{\alpha\beta} = \frac{1}{\Im\tau}
\begin{pmatrix}
|\tau|^2 & -\Re\tau \\ -\Re\tau & 1
\end{pmatrix},
\qquad \tau = C_0 + i e^{-\Phi},
\end{equation}
we can write the rescaling parameter as $\Omega = m_{22}^x(-m_{12}/m_{22})^z$.
To compute the effect of rescaling on the Ricci scalar we can use the standard
formula for the Weyl rescaling of the Ricci scalar (see for
example~\cite{Wald:1984rg}):
\begin{equation}
\tilde R = \Omega^{-2}\left[R - 18g^{\hat\mu\hat\nu}\nabla_{\hat\mu}\nabla_{\hat\nu}\log\Omega - 72g^{\hat\mu\hat\nu}(\nabla_{\hat\mu}\log\Omega)(\nabla_{\hat\nu}\log\Omega)\right].
\end{equation}

\paragraph{The $\lambda < 1$ case}
After the rescaling the Ricci scalar takes the following form:
\begin{equation}
\tilde R^{(\lambda < 1)} = \frac{\mathcal P^{(\lambda < 1)}(w_1, w_2)}{(1 - w_1^2 - w_2^2)^{\frac54 - x} w_2^{2+2z} (1 - w_1)^{2+2x-2z}} + \mathcal O(\chi),
\end{equation}
where $\mathcal P^{(\lambda < 1)}(w_1, w_2)$ is a polynomial in $w_1$ and $w_2$
with coefficients containing $x$, $z$, and $\lambda$. After a careful inspection
it is evident that there is no choice of $x$ and $z$ that removes the
denominator. One can also show that the numerator $\mathcal P^{(\lambda <
1)}(w_1, w_2)$ is non-zero for any choice of $x$, $z$, and $\lambda < 1$,
therefore the singularity in the curvature cannot be completely removed.

One can now study what type of singular behaviour the terms in the denominator
entail. The term $(1 - w_1^2 - w_2^2)$ is just the original ring singularity
along the circle $w_1^2 + w_2^2 = 1$. The term $w_2$ leads to singularities on
parts of the ring corresponding to $\theta\in\{0,\pi\}$ or $\phi=\pi/2$, while
the term $(1 - w_1)$ reduces the singularity to a single point $(w_1, w_2) = (1,
0)$, which is equivalent to the value $\theta = 0$. Thus we see that the least
singular behaviour that we can get is achieved by choosing $x \geq 5/4$ and $z
\leq -1$ which leads to a singularity of type $(1 - w_1)^a$ with $a \geq 9/4$,
\textit{i.e.} in this case we only have a singularity at a single point. It
would be interesting to understand the meaning of these frames.

\paragraph{The $\lambda=1$ case}
We can now repeat the same analysis for the case $\lambda = 1$. The transformed
Ricci scalar has the form
\begin{equation}
\tilde R^{(\lambda = 1)} = \left(\frac{8}{\chi^2} - 1\right)\frac{\mathcal P^{(\lambda =1)}(w_1, w_2)}{(4 - w_1^2 - 4w_2^2)^{\frac94 - x} w_2^{2z} (2 - w_1)^{2+2x-2z}} + \mathcal O(\chi).
\end{equation}
First of all, also in this case it can be shown that $\mathcal P^{(\lambda
=1)}(w_1, w_2)$ cannot be identically zero for any choice of $x$ and $z$. This
means that the singularity $\chi^{-2}$ in the radial coordinate can never be
removed. Notice however that the term $(2 - w_1)$ in the denominator is never
zero since $-1 \leq w_1 \leq 1$, and therefore we can arrange that the
singularity in the angular directions is removed completely by choosing $x \geq
9/4$ and $z \leq 0$.

%%%%%%%%%%%%%%%%%%%%%%%%%%%%%%%%%%%%%%%%%%%%%%%%%%%%
%
% Conclusions and Outlook
%
%%%%%%%%%%%%%%%%%%%%%%%%%%%%%%%%%%%%%%%%%%%%%%%%%%%%
\section{Conclusions and Outlook}%
\label{sec:conclusion}

In this paper we presented the uplift of the GPPZ solution to ten dimensions.
The original GPPZ solution involved two real scalars, which correspond to the
norm of a ${\cal N}=1$ supersymmetric mass deformation of ${\cal N}=4$ SYM
(which is a complex parameter) and the norm %of the source that couples to the
gaugino bilinear. The complex scalars corresponding to the sources that couple
to the complex QFT operators are part of a consistent truncation of $D=5$
maximal supergravity containing, besides the two complex scalars, the metric and
a $\textup{U}(1)$ gauge field, which is dual to the $\textup{U}(1)_R$ current.
We generalised the GPPZ solution to complex fields, which either have constant
phases or a spacetime dependent but $\textup{U}(1)$-covariantly constant phase.
In ten dimensions the solution with non-zero phases can be obtained from the one
with no phases by a combination of coordinate transformation on $S^5$ and an
$\textup{SO}(2)$ rotation in $\textup{SL}(2,R)$ that correspond to a
$\textup{U}(1)_R$ transformation and the bonus $\textup{U}(1)$ transformation on
the QFT side, respectively. The ten-dimensional solution has an infrared
singularity whose structure depends on the ratio of the mass deformation
parameter and the gaugino condensate, which we denoted by $\lambda$.
It was argued in~\cite{Girardello:1999bd}
%The study of the spectrum of glueballs in the five-dimensional solution confirm
%the argument of~\cite{Gubser:2000nd}
that the singularity is acceptable only if $\lambda \leq 1$. In the $\lambda <
1$ case, the non-compact part of the metric is regular and there is only a ring
singularity in the $S^5$ part. In the $\lambda=1$ case there is a singularity
both in the radial direction (but milder than that of the five dimensional
solution) and also at two points on $S^5$. Intriguingly, one can find conformal
frames such that there is only a singularity in the Ricci scalar at one point in
$S^5$ in the case of $\lambda<1$, or only in the radial coordinate in the case
of $\lambda=1$. We note however that these comments are based on the Ricci
scalar, and we have not checked the behaviour of other curvature invariants.

The metric and axion-dilaton field of the solution with zero phases agree
exactly with those of the a partial uplift by Pilch and Warner and the
singularity structure agrees with their findings. However we find additional
subleading terms in the near-singularity metric, which do not fit in the
interpretation of the singularity as due to seven-branes.

We also compared the asymptotics of the uplifted solution with that of the
Polchinski-Strassler solution~\cite{Polchinski:2000uf} and the zero temperature
limit of the Freedman-Minahan solution~\cite{Freedman:2000xb}. Our results agree
with that of Freedman-Minahan and disagree with Polchinski-Strassler. We have
checked that we use the same boundary condition with Polchinski-Strassler, {\it
i.e.} the non-normalizable modes (sources) are in exact agreement: the metric is
asymptotically $AdS_5 \times S^5$ and the part of the 3-form field strength that
depends on the deformation parameter are in exact agreement. The subleading
terms however disagree. Recalling that the subleading terms (up to the order
where the vevs appear) are uniquely fixed in terms of the sources by the field
equations~\cite{deHaro:2000vlm}, we infer that the Polchinski-Strassler $10d$
fields do not satisfy the IIB field equations near the boundary of AdS.

There are still many things to do. The most urgent question is whether the
solution in sourced by branes. While we have checked explicitly with Mathematica
that the uplift solution solves the IIB equations, there may be delta-function
sources in the field equations coming from the couplings of IIB fields to the
worldvolume of branes. In principle, one can detect such terms by a careful
analysis of the way the field equations are satisfied. To illustrate this point
suppose that while checking the field equations one needs to evaluate
\begin{equation} \label{delta}
\Box \frac{1}{r^{n-2}},
\end{equation}
where $\Box$ is the Laplacian in $n$ (Euclidean) dimensions and $r$ is the
distance from the origin. This term is proportional to a delta function and
hunting for such terms should provide the delta function sources. However, given
the complexity of the uplifted solution and of the IIB field equations and the
fact that we do not know the positions and orientations of the branes (if
present at all), this is a rather daunting task.

One way to proceed is to observe that one may check that~\eqref{delta} has
indeed a delta function by multiplying it with a smooth test function and
integrating over all space: if there is a source then the answer for the
integral will be non-zero, otherwise it will be zero. We have devised a version
of this method to check whether the uplifted solution is supported by branes or
not.
%Our preliminary results suggest that there are 5-brane and possibly 7-branes sources.
The analysis is complicated by the fact that the bulk solution has a (conformal)
boundary and it is singular, and thus one needs to carefully disentangle the
three possible contributions to the integral: due to the boundary, from the
singularity and due to delta functions. We will report on this analysis
elsewhere.

To understand the implications of the singularity it will also be useful to
analyze how different probes respond when placed in this background. Given that
the solution is very explicit it should be straightforward to perform such
analysis.

It is also interesting to understand to which of the many $\mathcal{N}=1^*$
vacua the GPPZ flow and the uplifted solution correspond to. Now that we have
the solution in ten dimensions one may proceed to extract the 1-point functions
of operators other than the gaugino bilinear. This can be done following the
framework of Kaluza-Klein holography~\cite{Skenderis:2006uy} and it should allow
us to definitely establish whether this solution is dual to the confining vacuum
of $\mathcal N=1^*$, a different vacuum of this theory, or that it is unphysical
(because of the singularity) and it should be discarded. We hope to return to
this and related issues in the near future.

Finally, it is natural to apply the uplift techniques to other five-dimensional
solutions, as, for instance, the solution of $\mathcal{N}=8$ gauge supergravity
dual to the $\mathcal{N}=2^*$ theory, a deformation of $\mathcal{N}=4$ SYM by a
mass for two of the chiral multiplets~\cite{Gubser:2000nd, Pilch:2000ej,
Pilch:2000ue, Brandhuber:2000ct}. Recently the approach of consistent truncation
has been used to compute the ten-dimensional solution dual to $\mathcal{N}=2^*$
on $S^4$~\cite{Bobev:2018hbq}. It would be interesting to study other cases,
with less supersymmetry or different compact manifolds.

\paragraph{Note Added:}
Soon after this paper appeared in the arXiv the paper~\cite{Bobev:2018eer}
appeared, which has overlap with our results. The two solutions are the same expressed in different coordinates.
We thank the authors of ~\cite{Bobev:2018eer} for communication regarding the comparison of the solutions.

%%%%%%%%%%%%%%%%%%%%%%%%%%%%%%%%%%%%%%%%%%%%%%%%%%%%
%
% Acknowledgments
%
%%%%%%%%%%%%%%%%%%%%%%%%%%%%%%%%%%%%%%%%%%%%%%%%%%%%
\acknowledgments{}
We acknowledge the hospitality of the Galileo Galilei Institute for Theoretical
Physics, where this work was initiated during the workshop ``Supergravity: what
next?''.
KS is supported in part by the Science and Technology Facilities Council
Consolidated Grant ``New Frontiers in Particle Physics and Cosmology'',
ST/P000711/1. This project has received funding from the European Union's
Horizon 2020 research and innovation programme under the Marie Sklodowska-Curie
grant agreement No 690575.
SS acknowledges the funding from the Mayflower Scholarship from the University
of Southampton.
MP is partly supported by ILP LABEX (ANR-10-LABX-63) and the Idex SUPER
(ANR-11-IDEX-0004-02).

%%%%%%%%%%%%%%%%%%%%%%%%%%%%%%%%%%%%%%%%%%%%%%%%%%%%
%
% APPENDIX
%
%%%%%%%%%%%%%%%%%%%%%%%%%%%%%%%%%%%%%%%%%%%%%%%%%%%%
\appendix

%%%%%%%%%%%%%%%%%%%%%%%%%%%%%%%%%%%%%%%%%%%%%%%%%%%%
%
% IIB uplift of D=5 maximal supergravity
%
%%%%%%%%%%%%%%%%%%%%%%%%%%%%%%%%%%%%%%%%%%%%%%%%%%%%
\section{IIB uplift of \texorpdfstring{$D=5$}{D=5} maximal supergravity}%
\label{app:IIb_uplift}

$D=5$ maximal $\textup{SO}(6)$ gauged
supergravity~\cite{Gunaydin:1984qu,Pernici:1985ju,Gunaydin:1985cu} is a
consistent truncation of IIB supergravity around AdS$_5\times S^5$\,. Its field
content comprises the $D=5$ metric $g_{\mu\nu}$ together with 42 scalar fields
parametrizing a $27\times27$ symmetric ${\rm E}_{6(6)}$ matrix which we
parametrize in an ${\rm SL}(6)\times{\rm SL}(2)$ basis as
\begin{eqnarray}
M_{MN} &=& \left(
\begin{array}{cc}
{M}_{ab,cd}&{M}_{ab}{}^{c\beta}\\
{M}^{a\alpha}{}_{cd} & M^{a\alpha,c\beta}
\end{array}
\right)
\;,
\quad
\mbox{with inverse}\;\;
M^{MN} ~=~ \left(
\begin{array}{cc}
{M}^{ab,cd}&{M}^{ab}{}_{c\beta}\\
{M}_{a\alpha}{}^{cd} & M_{a\alpha,c\beta}
\end{array}
\right)
\;,%
\label{MD5}
\end{eqnarray}
according to the decomposition of the fundamental representation of ${\rm
E}_{6(6)}$ as
\begin{eqnarray}
{\bf 27}\longrightarrow ({\bf 15},{\bf 1}) \oplus ({\bf 6'},{\bf 2})
\;,
\end{eqnarray}
under ${\rm SL}(6)\times{\rm SL}(2)$. Indices $a, b, c, d=1, \dots, 6$, and
$\alpha, \beta=1,2$, label the fundamental representations of ${\rm SL}(6)$ and
${\rm SL}(2)$, respectively. Index pairs $ab$ and $cd$ in~\eqref{MD5} are
antisymmetric. The remaining bosonic field content in five dimensions is given
by 15 non-abelian vectors fields $A_\mu^{ab}$ and 12 topologically massive
two-forms $B_{\mu\nu,a\alpha}$. The truncation we are eventually interested in
and which carries the GPPZ solution~\cite{Girardello:1999bd} carries four scalar
fields, a single vector and no two-forms.

In this section we collect the relevant IIB uplift formulae of $D=5$
supergravity from~\cite{Baguet:2015sma} (see also~\cite{Khavaev:1998fb,
Nastase:2000tu, Pilch:2000ue, Lee:2014mla, Ciceri:2014wya}), in the next section
we explicitly evaluate these formulas for the four-scalar truncation. The IIB
fields are expressed in terms of the $D=5$ fields introduced above while their
dependence on the five internal coordinates $y^m$ is carried by the fundamental
$S^5$ sphere harmonics, ${\cal Y}^a$, ($a=1, \dots, 6$), with ${\cal Y}^a{\cal
Y}^a=1$, and the $S^5$ Killing vectors
\begin{eqnarray}
 {\cal K}_{[ab]\,m} &=& -\sqrt{2}\, {\cal Y}^{[a}\partial_m{\cal Y}^{b]}
 \;,
 \qquad
 m=1, \dots, 5\;.
 \label{KV}
\end{eqnarray}
By $\mathring{G}_{mn}$ we denote the round metric on $S^5$ which can be
expressed as
\begin{eqnarray}
\mathring{G}_{mn} &=&
{\cal K}_{[ab]\,m} {\cal K}_{[ab]\,n}
\;,
\label{G0}
\end{eqnarray}
in terms of the Killing vectors~\eqref{KV}. We also define its volume form
\begin{eqnarray}%
\label{eq:S5_volume_form}
\mathring{\omega}_{klmnp}~\equiv~\sqrt{\det\mathring{G}\,}\,\varepsilon_{klmnp}
~\equiv~5\,\partial_{[k}\mathring{C}{}_{lmnp]}
\;,
\label{volume}
\end{eqnarray}
in terms of a 4-form potential $\mathring{C}_{klmn}$\,. We will also need the
tensors
\begin{eqnarray}
{\cal K}_{[ab]\,mn} &\equiv&
\partial_m {\cal K}_{[ab]\,n}-\partial_n {\cal K}_{[ab]\,m}
\;,
\nonumber\\
{\cal K}_{[ab]\,klm} &\equiv&
\frac12\,\mathring{\omega}_{klmnp}\,{\cal K}_{[ab]}{}^{np}
\;,
\end{eqnarray}
where indices $n, p$ on the r.h.s.\ are raised with the background
metric~\eqref{G0}.

In terms of these objects, the IIB metric takes the following form
\begin{eqnarray}
ds^2 &=&
 \Delta^{-2/3}(x,y)\,{g}_{\mu\nu}(x) \, dx^\mu dx^\nu
\nonumber\\
 &&{}
 +
G_{mn}(x,y)
\left( dy^m + {\cal K}_{[ab]}{}^m(y) A_\mu^{ab}(x) dx^\mu\right)\left( dy^n +{\cal K}_{[cd]}{}^n(y) A_\nu^{cd}(x)  dx^\nu\right)
 \;,
 \label{metricIIB}
\end{eqnarray}
with the internal block $G_{mn}(x,y)$ given by inverting the matrix
\begin{eqnarray}
 \mathcal G^{mn}(x,y) &=& \,\Delta(x,y)^{2/3}\, {\cal K}_{[ab]}{}^m(y) {\cal K}_{[cd]}{}^n(y)\,M^{ab,cd}(x)
\;,
\end{eqnarray}
in terms of the submatrix $M^{ab,cd}(x)$ from~\eqref{MD5}. We use indices $\mu,
\nu$ and $m, n$ for the external five and internal five coordinates,
respectively. The warp factor $\Delta(x,y)$ is defined as
\begin{eqnarray}
\Delta(x,y) &=& \frac{(\det G_{mn}(x,y))^{1/2}}{(\det\mathring{G}_{mn}(y))^{1/2}}
\;.
\end{eqnarray}
The IIB dilaton and axion combine into a symmetric ${\rm SL}(2)$ matrix $m_{\alpha\beta}$
whose inverse is given by
\begin{eqnarray}
m^{\alpha\beta}(x,y) &=&
 \Delta(x,y)^{4/3}\, {\cal Y}^a(y) {\cal Y}^b(y) \, M^{a\alpha,b\beta}(x)
\;.
\label{DA}
\end{eqnarray}
The relevant components of the IIB 2-form doublet and 4-form gauge potentials
are given by
\begin{eqnarray}
C_{mn}{}^\alpha
 &=&
-\frac12\,
\varepsilon^{\alpha\beta} \Delta^{4/3}\, m_{\beta\gamma}\, {\cal Y}^c\,{\cal K}_{[ab]\,mn}
\,{M}_{ab}{}^{c\gamma}
\;, \label{C2C4}\\
C_{\mu kmn}&=&
-\frac18\,\mathring{\omega}_{kmnpq} \mathring{\nabla}^p {\cal K}_{[ab]}{}^{q} \,A_\mu{}^{ab}
-{\cal K}_{[ab]}{}^p A_\mu{}^{[ab]}\,\mathring{C}_{pkmn}
\;,\nonumber\\
 C_{m\,\mu\nu\rho} &=&
-\frac1{32}\,{\cal K}_{[ab]\,m}\left(
2\,\sqrt{|{ g}|} \,\varepsilon_{\mu\nu\rho\sigma\tau}\,M_{ab,cd} F^{\sigma\tau\,cd}
+3\,\sqrt{2}\,\varepsilon_{abcdef}\,
\partial_{[\mu} A_{\nu}{}^{cd} A_{\rho]}{}^{ef} \right)
\;,\nonumber\\
C_{klmn} &=& \mathring{C}_{klmn} -
\frac{1}{6}\,\mathring\omega_{klmnp}\,  \mathring{G}^{pq} \, \Delta^{-1} \partial_q \Delta
 \;,
 \nonumber\\
 C_{\mu\nu\rho\sigma} &=&
-\frac1{16}\, {\cal Y}^{a} {\cal Y}^{b}
\,
\sqrt{|{g}|}\,\varepsilon_{\mu\nu\rho\sigma\tau}
{D}^\tau {M}_{bc,N} {M}^{N\,ca}
+ \Lambda_{\mu\nu\rho\sigma}
\;,
\nonumber
\end{eqnarray}
with $F_{\mu\nu}{}^{ab}$ the five-dimensional field strength of
$A_\mu{}^{ab}$\,.
The function $\Lambda_{\mu\nu\rho\sigma}(x)$ in the last line is defined by
integrating
\begin{eqnarray}
D_{[\mu} \Lambda_{\nu\rho\sigma\tau]}
&=&
\frac1{600}\,\sqrt{|{ g}|}\,\varepsilon_{\mu\nu\rho\sigma\tau}
 \left(
10\, \delta_h^d\delta^a_e
+  2\,{M}^{fd,ga} {M}_{gh,fe} -
{M}_{e\alpha}{}^{ga}
 {M}_{gh}{}^{d\alpha}
%+ {\cal X}^{(ab)ec,d}{}_e
\right) M^{bh,ec}\,\delta_{cd} \delta_{ab}
\nonumber\\
&&{}
-\frac1{480}\,
\,\sqrt{|{ g}|}\,\varepsilon_{\mu\nu\rho\sigma\tau} D_{\lambda}
\left({M}^{N\,ac} \,{D}^\lambda {M}_{ac,N} \right)
\label{defLambda}\\
&&{}
+\frac{1}{240}\,
\sqrt{|{ g}|}\,\varepsilon_{\mu\nu\rho\sigma\tau}\,
{M}_{ab,cd} \, {F}_{\kappa\lambda}{}^{ab}\,{F}^{\kappa\lambda\,cd}
+\frac1{32}\,\sqrt{2}\,
 \,\varepsilon_{abcdef}\, {F}_{[\mu\nu}{}^{ab} F_{\rho\sigma}{}^{cd} A_{\tau]}{}^{ef}
\;.
\nonumber
\end{eqnarray}
The $p$-forms~\eqref{C2C4} are given in the standard Kaluza-Klein basis
\begin{eqnarray}
Dy^m&=&dy^m + {\cal K}_{[ab]}{}^m(y) A_\mu^{ab}(x) dx^\mu
\;,
\label{DY}
\end{eqnarray}
c.f.~\eqref{metricIIB}. As compared to the full uplift
formulas~\cite{Baguet:2015sma} we have suppressed in the $p$-forms~\eqref{C2C4}
all terms anti-symmetric in more than one vector field since these will not
survive in the truncation to a single vector field which is of interest here.

%%%%%%%%%%%%%%%%%%%%%%%%%%%%%%%%%%%%%%%%%%%%%%%%%%%%
%
% Parametrization of the scalar matrix
%
%%%%%%%%%%%%%%%%%%%%%%%%%%%%%%%%%%%%%%%%%%%%%%%%%%%%
\section{Parametrization of the scalar matrix}%
\label{app:E6}

In this appendix we spell out the explicit parametrization of the scalar
E$_{6(6)}$ matrix $M_{MN}$~\eqref{MD5} in the 4-scalar truncation of $D=5$
maximal supergravity. To this end, we go to a complex basis, in which the ${\rm
SL}(6)$ vector decomposes according to
\begin{eqnarray}
\{X_a\} &\longrightarrow&
\{ X_i, X_{\overline{i}}=X_i^* \}
\;,\qquad
i, \overline{i} = 1, 2, 3
\;.
\end{eqnarray}
In this decomposition, the E$_{6(6)}$ matrix~\eqref{MD5} decomposes as
\begin{eqnarray}
M_{MN} &=&
 \left(
\begin{array}{ccccc}
{M}_{ij,kl}&{M}_{ij,k\overline{l}}&{M}_{ij,\overline{kl}}&{M}_{ij}{}^{k\beta}&{M}_{ij}{}^{\overline{k}\beta}\\
{M}_{i\overline{j},kl}&{M}_{i\overline{j},k\overline{l}}&{M}_{i\overline{j},\overline{kl}}
&{M}_{i\overline{j}}{}^{k\beta}&{M}_{i\overline{j}}{}^{\overline{k}\beta}\\
{M}_{\overline{ij},kl}&{M}_{\overline{ij},k\overline{l}}&{M}_{\overline{ij},\overline{kl}}
&{M}_{\overline{ij}}{}^{k\beta}&{M}_{\overline{ij}}{}^{\overline{k}\beta}\\
{M}^{i\alpha}{}_{kl}&{M}^{i\alpha}{}_{k\overline{l}}&{M}^{i\alpha}{}_{\overline{kl}}&{M}^{i\alpha,k\beta}
&{M}^{i\alpha,\overline{k}\beta}\\
{M}^{\overline{i}\alpha}{}_{kl}&{M}^{\overline{i}\alpha}{}_{k\overline{l}}
&{M}^{\overline{i}\alpha}{}_{\overline{kl}}&{M}^{\overline{i}\alpha,k\beta}
&{M}^{\overline{i}\alpha,\overline{k}\beta}\\
\end{array}
\right)
\;,%
\label{MSO3}
\end{eqnarray}
with its non-vanishing entries given in terms of the ${\rm SL}(2)$ vector
$v^\alpha ~\equiv~ (1,-i)^\alpha$ by
\begin{eqnarray}
{M}_{ij,kl} &=& -\delta_{ij}{}^{kl}\,e^{-i\,(\varphi+\omega)}\,
\sinh \left(\frac{2 m}{\sqrt{3}}\right)\sinh \left(2 \sigma \right)
\;,\nonumber\\
{M}_{ij,\overline{kl}} &=& \delta_{ij}{}^{\overline{kl}}\,
\cosh \left(\frac{2 m}{\sqrt{3}}\right)\cosh \left(2 \sigma \right)
\;,
\nonumber\\
M_{ij}{}^{k\alpha} &=&
\frac12\,\varepsilon^{ijk}\,e^{-i\varphi}\,\sinh \left(\frac{2 m}{\sqrt{3}}\right)\cosh \left(2 \sigma \right)
\,v^\alpha
\;,
\nonumber\\
M_{\overline{ij}}{}^{{k}\alpha} &=&
-\frac12\,\varepsilon^{ijk}\,e^{i\omega}\,
\cosh \left(\frac{2 m}{\sqrt{3}}\right)\sinh \left(2 \sigma \right)
\,v^\alpha
\;,
\nonumber\\
M_{i\overline{j}}{}^{\overline{k}\alpha} &=&
\frac14\,\varepsilon^{ijk}\,e^{-i\varphi}\,\sinh \left(\frac{4 m}{\sqrt{3}}\right)
\,v^\alpha\;,
\nonumber\\
M^{i\alpha,{j}\beta} &=&
\frac12\,e^{2i\varphi}\,\sinh^2 \left(\frac{2 m}{\sqrt{3}}\right)
(v^\alpha v^\beta)^*
-\frac12\,e^{i(\omega-\varphi)}\,\sinh \left(\frac{2 m}{\sqrt{3}}\right)
\sinh \left(2 \sigma \right)
v^\alpha v^\beta
\;,
\nonumber\\
M^{i\alpha,\overline{j}\beta} &=&
\frac12\,\cosh^2 \left(\frac{2 m}{\sqrt{3}}\right)
(v^\alpha)^*\,v^\beta
+\frac12\,\cosh \left(\frac{2 m}{\sqrt{3}}\right)
\cosh \left(2 \sigma \right)
v^\alpha\,(v^\beta)^*
\;,
\end{eqnarray}
together with those components related by complex conjugation. Plugging this
explicit form of the scalar matrix into the uplift formulas of
Appendix~\ref{app:IIb_uplift} yields the IIB uplift of the 4-scalar truncation
of the $D=5$ theory which we describe in Section~\ref{sec:uplift_gppz}.

%%%%%%%%%%%%%%%%%%%%%%%%%%%%%%%%%%%%%%%%%%%%%%%%%%%%
%
% Uplift in Pilch-Warner Coordinates
%
%%%%%%%%%%%%%%%%%%%%%%%%%%%%%%%%%%%%%%%%%%%%%%%%%%%%
\section{Uplift in Pilch-Warner Coordinates}%
\label{app:uplift_in_pw_coords}

In this appendix we present the uplift solution in the coordinates introduced
in~\cite{Pilch:2000fu}. First let us recall the definition of the new radial
coordinate $t$ and other constants
\begin{equation}
t = e^{-(y-C_1)}, \qquad \lambda = e^{3(C_2 - C_1)}, \qquad C_1 = \log\left(\frac{m_0}{\sqrt3}\right), \qquad C_2 = \frac13\log\left(\frac{\sigma_0}{2}\right).
\end{equation}
where $C_1$ and $C_2$ are the $5d$ integration constants and $m_0, \sigma_0$ are
related to the leading asymptotic behaviour of the $5d$ fields $m$ and $\sigma$
(note that $m_0, \sigma_0$ differ by constants relative to the ones
in~\cite{Pilch:2000fu}). Defining $\mu, \nu$ as in~\eqref{eq:def_mu_nu}, the
solution of the first order equation in terms of these variables takes the form
\begin{align}
\mu(t) &= \sqrt{\frac{1+\lambda t^3}{1-\lambda t^3}}, \qquad \nu(t) = \sqrt{\frac{1+t}{1-t}}.
\end{align}
Yet another definition of the radial coordinate introduced
in~\cite{Pilch:2000fu} is the following
\begin{equation}
\chi = 2(1-t)^{1/2}.
\end{equation}

For the internal manifold we think of the round sphere as embedded in
$\mathbb{R}^6$ described by the coordinates $y_1,\dots y_6$, so that on the
sphere $\vec y^{\,2} = 1$. The six coordinates can be thought of as split into
two triplets
\begin{equation}
\vec y = (y_1,\dots y_6) \to (u_1, u_2, u_3, v_1, v_2, v_3) \equiv (\vec u, \vec v).
\end{equation}
The diagonal $\textup{SO}(3)$ acts on $u^i$ and $v^j$ simultaneously in the
vector representation. In~\cite{Pilch:2000fu} the authors show that $\vec u$ and
$\vec v$ can be written as
\begin{align}
\vec u &= R\begin{pmatrix}0 \\ 0 \\ \cos\theta \end{pmatrix} \equiv R.\vec u_0 \\
\vec v &= R\begin{pmatrix}0 \\ \sin\theta \sin\phi \\ \sin\theta \cos\phi \end{pmatrix} \equiv R.\vec v_0
\end{align}
with $\theta\in[0,\pi/2]$, $\phi\in[0,\pi]$ and $R =
R(\alpha_1,\alpha_2,\alpha_3)$ a generic $\textup{SO}(3)$ matrix parametrized by
three Euler angles $\alpha_i$. Following~\cite{Pilch:2000fu} we further define
\begin{alignat}{2}
w_1 &= 2\, \vec u^{\,2} - 1 &&= \cos(2\theta) \\
w_2 &= 2\, \vec u.\vec v &&= \sin(2\theta) \cos \phi
\end{alignat}
such that the internal manifold is described by the coordinates
$\{\alpha_1,\alpha_2,\alpha_3,w_1,w_2\}$.

To write form-fields in terms of these coordinates it is useful to compute the
differentials $du^i$ and $dv^i$. In the coordinates just introduced this
translates to
\begin{equation}
d\vec u = dR.\vec u_0 + R.d\vec u_0 = R(R^{-1}dR.\vec u_0 + d\vec u_0).
\end{equation}
The Maurer-Cartan form $R^{-1}dR$ can be decomposed into left-invariant 1-forms
$\sigma^i$, which we list below in~\eqref{LIF} so that the differential $du^i$
can be written as
\begin{equation}
d\vec u = R(i\sigma^i T^i.\vec u_0 + d\vec u_0),
\end{equation}
and analogously for $dv^i$. Since in the quantities we are interested in the
$\textup{SO}(3)$ indices are always contracted the overall factor of $R$ in the
differentials drops out. Thus (using the fact that $(T^j)^{ik} =
i\epsilon^{ijk}$ for $\textup{SO}(3)$) in $\textup{SO}(3)$-invariant expressions
we can substitute
\begin{align}
du^i &\to \epsilon^{ijk}u_0^j\sigma^k + du_0^i \\
dv^i &\to \epsilon^{ijk}v_0^j\sigma^k + dv_0^i.
\end{align}

We take  the $\textup{SO}(3)$ rotation matrix as   $R(\alpha_1, \alpha_2, \alpha_3) =  e^{ i \alpha_3 T_3} e^{ i \alpha_2 T_1} e^{ i \alpha_1 T_3}$. Then $R^{-1} d R = i  \sum_{i=1}^3 \sigma_i T_i$
gives the left invariant forms 
\begin{equation}
\label{LIF}
\begin{split}
\sigma^1 &= \cos\alpha_1 \; d\alpha_2 + \sin\alpha_1 \sin\alpha_2 \; d\alpha_3 \\
\sigma^2 &= \sin\alpha_1 \; d\alpha_2 - \cos\alpha_1 \sin\alpha_2 \; d\alpha_3 \\
\sigma^3 &= d\alpha_1 + \cos\alpha_2 \; d\alpha_3.
\end{split}
\end{equation}
The ranges for the Euler angles are $\alpha_1, \alpha_3 \in [0,2\pi]$, and
$\alpha_2 \in[0,\pi]$. One may check directly that these forms indeed satisfy
\begin{equation}
d\sigma^a = \frac12\epsilon^{abc}\sigma^b\wedge\sigma^c.
\end{equation}
The following variables (also introduced in~\cite{Pilch:2000fu}) are also
useful,
\begin{align}
\zeta &= 1 - w_1^2 - w_2^2 \\
\hat \Omega &= \frac13\sqrt{4(1-w_1^2-w_2^2)+3w_1^2} =\frac13\sqrt{4-w_1^2-4w_2^2},
\end{align}
These variables are the coefficients of the leading terms of the warp factor in
the $t \to 1 $ and $\lambda \to 1$, $t \to 1$ limits.

We are now ready to present the solution in terms of these variables.

\paragraph{Warp-Factor}
The warp-factor $\Delta$ we used earlier and the warp factor $\xi$
in~\cite{Pilch:2000fu} are related by
\begin{equation}
\xi^2 = \Delta^{-8/3}.
\end{equation}
In the new coordinates we find
\begin{equation}
\begin{split}
\xi^2 &= \frac{1}{\left(1-t^2\right)^4 \left(1-\lambda ^2 t^6\right)^2} \times \\
 &\Biggl[\left(1+t^2\right)^2 \left(1-\lambda ^2 t^8\right)^2 - 4 w_1^2\; t^4 \left(1-\lambda t^2\right)^2 \left(1 + \lambda
  t^4\right)^2 - 4 w_2^2\; t^4 \left(1 + \lambda t^2\right)^2 \left(1-\lambda t^4\right)^2
\Biggr]
\end{split}
\end{equation}

\paragraph{Metric}
The uplifted ten-dimensional metric was already obtained earlier by Pilch and
Warner~\cite{Pilch:2000fu}, see equations (6.1)--(6.7) in their text. It takes a
block form containing the $AdS$ and the $S^5$ parts as follows
\begin{align}
ds_{10}^2 &= \xi^{1/2}ds^2_{1,4} + \xi^{-3/2}ds_5^2 \\
ds^2_{1,4} &= e^{2\phi(y)}\eta_{\mu\nu}dx^\mu dx^\nu + dy^2 \\
\begin{split}
ds_5^2 &= a_1 du^i du^i + 2a_2 du^i dv^i + a_3 dv^i dv^i \\
&\quad{} + a_4(u^i dv^i + v^i du^i)^2 + 2a_5 (u^i dv^i) (v^j du^j) + 2 a_6 (u^i du^i) (v^j dv^j).
\end{split}
\end{align}
The coefficients $a_i$ of the internal metric can be found in equation~(6.3) in
the Pilch and Warner text~\cite{Pilch:2000fu}. We can expand the fields $\mu(t)$
and $\nu(t)$ in terms of the radial coordinate $t$ to get the following
expressions for the coefficients $a_i$:
\begin{align}
a_1 &= \frac{\left(1+\lambda t^4\right) \left(1+t^2 (1-2 w_1) \left(1-\lambda t^2\right)-\lambda t^6\right)}{\left(1-t^2\right)^2 \left(1-\lambda ^2 t^6\right)} \\
a_2 &= \frac{-2 w_2 t^2 \left(1+\lambda t^2\right) \left(1-\lambda t^4\right)}{\left(1-t^2\right)^2 \left(1-\lambda ^2 t^6\right)} \\
a_3 &= \frac{\left(1+\lambda t^4\right) \left(1+t^2 (1+2 w_1) \left(1-\lambda t^2\right)-\lambda t^6\right)}{\left(1-t^2\right)^2 \left(1-\lambda ^2 t^6\right)} \\
a_4 &= \frac{t^2 \left(1+\lambda t^2\right)^2 \left(1+\lambda t^4\right) \left(1+3 t^2 \left(1-\lambda t^2\right)-\lambda t^6\right)}{\left(1-t^2\right)^3 \left(1-\lambda ^2 t^6\right)^2}\\
a_5 &= \frac{2 t^2 \left(1-\lambda ^2 t^4 \left(1+t^4 \left(1-\lambda ^2 t^4\right)\right)\right)}{\left(1-t^2\right)^2 \left(1-\lambda ^2 t^6\right)^2} \\
a_6 &= \frac{-4 t^2 \left(1+t^2\right) \left(1-\lambda t^2 \left(1+t^2 \left(1-\lambda t^2\right)\right)\right)}{\left(1-t^2\right)^3 \left(1-\lambda ^2 t^6\right)}.
\end{align}

\paragraph{Axion/Dilaton}
The axion/dilaton matrix $m_{\alpha\beta}$ is given by
\begin{equation}
m_{\alpha\beta} = \frac{1}{\xi}m_{ab} = \frac{1}{\xi}\begin{pmatrix}m_{11} & m_{12} \\ m_{12} & m_{22}\end{pmatrix}
\end{equation}
with the components
\begin{align}
m_{11} &= \frac{\left(1+\lambda t^4\right) }{\left(1-t^2\right)^2 \left(1-\lambda ^2 t^6\right)}
\left[(1-\lambda t^6) + t^2 (1 + 2 w_1) - \lambda t^4 (1 + 2 w_1)\right] \\
m_{22} &= \frac{\left(1+\lambda t^4\right) }{\left(1-t^2\right)^2 \left(1-\lambda ^2 t^6\right)}
\left[(1-\lambda t^6) + t^2 (1 - 2 w_1) - \lambda t^4 (1 - 2 w_1)\right] \\
m_{12} &= \frac{2 w_2 t^2 \left(1-\lambda t^4\right) \left(1+\lambda t^2\right)}{\left(1-t^2\right)^2 \left(1-\lambda ^2 t^6\right)}
\end{align}

\paragraph{2-Form Potential}
The 2-form potential is given by
\begin{equation}
C_\alpha = C_{mn \; \alpha}dy^m \wedge dy^n .
\end{equation}
The new basis for the 2-forms will be given by the following 6 two-forms
\begin{equation}
\{ dw_1, dw_2, \sigma_1 \}\wedge \{\sigma_2, \sigma_3 \}
\end{equation}
The expression for $C_\alpha$ in terms of this basis is rather complicated, but
reduces to a manageable expression in the $t \to 1 $ and $\lambda <1$ or $t \to
1$ limits, which we reported earlier in Section~\ref{sec:singularity}.

\paragraph{4-Form Potential}
The 4-form potential is given by
\begin{equation}
C = \mathring{C} + \frac{1}{4! \; \xi^2}(f_1 d_4^1 + f_2 d_4^2 + f_3 d_4^3)
\end{equation}
with the coefficients
\begin{align}
f_1 &= \frac{-12 w_2 \; t^4(1+\lambda t^2)^2 (1-\lambda t^4)^2}{(1-t^2)^4 (1-\lambda^2 t^6)^2} \\
f_2 &= \frac{12 w_1 \; t^4 (1-\lambda t^2)^2 (1+\lambda t^4)^2}{(1-t^2)^4 (1-\lambda ^2 t^6)^2} \\
f_3 &= \frac{48w_1 w_2 \; \lambda t^6 }{(1-t^2)^3 (1-\lambda^2 t^6)}
\end{align}
and the 4-forms
\begin{align}
d_4^1
&= \epsilon^{ijm}\epsilon^{kln}(u^m u^n - v^m v^n) du^i \wedge du^j \wedge dv^k \wedge dv^l \\
&= \frac{\zeta^{1/2}}{4}\frac{(w_1-1)}{(w_1+1)}\;dw_1\wedge dw_2\wedge\sigma^2\wedge\sigma^3 \nonumber \\
&\qquad{}+\frac{1}{2} (1+w_1^2-w_2^2)\;dw_1\wedge\sigma^1\wedge\sigma^2\wedge\sigma^3
+w_1 w_2\; dw_2\wedge\sigma^1\wedge\sigma^2\wedge\sigma^3 \nonumber \\
d_4^2
&= \epsilon^{ijm}\epsilon^{kln}(u^m v^n + v^m u^n) du^i \wedge du^j \wedge dv^k \wedge dv^l \\
&= \frac{\zeta^{1/2}}{4}\frac{w_2}{(w_1+1)}\;dw_1\wedge dw_2\wedge\sigma^2\wedge\sigma^3 \nonumber \\
&\qquad{} + w_1 w_2\; dw_1\wedge\sigma^1\wedge \sigma^2\wedge \sigma^3 + \frac{1}{2} (1-w_1^2+w_2^2)\; dw_2\wedge \sigma^1\wedge \sigma^2\wedge \sigma^3 \nonumber \\
d_4^3
&= du^i \wedge du^j \wedge dv^i \wedge dv^j \\
&= \frac{\zeta^{1/2}}{4}\frac{1}{w_1+1}\;dw_1\wedge dw_2\wedge\sigma^2\wedge\sigma^3 \nonumber \\
&\qquad{}+\frac{w_1}{2}\;dw_1\wedge\sigma^1\wedge\sigma^2\wedge\sigma^3+\frac{w_2}{2}\;dw_2\wedge\sigma^1\wedge\sigma^2\wedge\sigma^3 \nonumber
\end{align}

The combined 4-form reads
\begin{equation}
\begin{split}
C = \mathring{C} + \frac{t^4}{4\xi^2 (1-t^2)^4 (1-\lambda^2 t^6)^2}\Bigl[
&\phantom{{}+{}}\tilde{f}_1 \; dw_1 \wedge dw_2 \wedge \sigma^2 \wedge \sigma^3 \\
&{}+\tilde{f}_2 \; dw_1 \wedge \sigma^1 \wedge \sigma^2 \wedge \sigma^3 \\
&{}+\tilde{f}_3 \; dw_2 \wedge \sigma^1 \wedge \sigma^2 \wedge \sigma^3 \;\Bigr]
\end{split}
\end{equation}
with
\begin{align}
\tilde{f}_1 &= \frac{\zeta^{1/2}}{2}\frac{w_2}{1+w_1}(1+\lambda t^2)^2 (1-\lambda t^4)^2 \\
\tilde{f}_2 &= -w_2 \left[(1+\lambda t^2)^2 (1-\lambda t^4)^2 - w_1^2 (1-\lambda t^2)^2 (1+\lambda t^4)^2 - w_2^2 (1+\lambda t^2)^2 (1-\lambda t^4)^2\right] \nonumber \\
\tilde{f}_3 &= w_1 \left[(1-\lambda t^2)^2 (1+\lambda t^4)^2 - w_1^2 (1-\lambda t^2)^2 (1+\lambda t^4)^2 - w_2^2 (1+\lambda t^2)^2 (1-\lambda t^4)^2\right]. \nonumber
\end{align}

\paragraph{6-Form Potential}
The non-vanishing components of the 6-form potential are the following
\begin{align}
C_{\mu\nu\rho\sigma\tau\,m}{}^\alpha &=
\omega_{ \mu\nu\rho\sigma\tau}\,\Xi_{m}{}^\alpha\;,
\nonumber\\
C_{\mu\nu\rho\sigma,\,mn}{}^\alpha &= \omega_{ \mu\nu\rho\sigma\tau}\,g^{\tau\lambda}
\Xi_{\lambda\,mn}{}^\alpha\;,
\end{align}
We can transform the one-forms $\Xi^\alpha$ and two-forms $\Xi_\lambda^\alpha$
into the Pilch-Warner basis and re-write some of the differentials in terms of
vector harmonics. The result for the one-forms is
\begin{equation}
\Xi^\alpha = \frac{\sqrt3}{2}
\frac{\left(\lambda t^3+t\right) \left(\lambda ^3 t^{12}+\lambda ^3 t^{10}+\lambda ^2 t^8-3 \lambda (\lambda +1) t^6+\lambda t^4+t^2+1\right)}{\left(t^2-1\right)^2 \left(\lambda ^2 t^6-1\right)^2}
\begin{pmatrix}
Y_v^{(1,1)} \\ -Y_v^{(1,0)}
\end{pmatrix}.
\end{equation}
For the two-forms we get
\begin{equation}
\begin{split}
\Xi_y^\alpha &=
\frac{\sqrt3}{\zeta}\frac{t \left(\lambda t^2-1\right) \left(\lambda t^4+1\right)}{\left(t^2-1\right) \left(\lambda ^2 t^6-1\right)}
\begin{pmatrix}
(1+w_1) \sqrt{\zeta} \sigma^1 \\
-\frac{w_1 w_2}{1+w_1}dw_1 - (1-w_1)dw_2 + (1-w_1)\sqrt{\zeta}\sigma^1
\end{pmatrix}
\wedge
\begin{pmatrix}
Y_v^{(1,0)} \\ Y_v^{(1,1)}
\end{pmatrix}
\\
&{} + \frac{\sqrt3}{\zeta}\frac{t(\lambda ^2 t^6+3 \lambda t^4-3 \lambda t^2-1)}{\left(t^2-1\right) \left(\lambda ^2 t^6-1\right)}
\begin{pmatrix}
-\frac12\frac{\zeta}{1+w_1}dw_1 + w_2\sqrt{\zeta}\sigma^1 - w_1 dw_1 - w_2 dw_2 \\
-\frac12\frac{\zeta}{1+w_1}dw_1 + w_2\sqrt{\zeta}\sigma^1
\end{pmatrix}
\wedge
\begin{pmatrix}
Y_v^{(1,1)} \\ Y_v^{(1,0)}
\end{pmatrix}
\end{split}
\end{equation}
Note that there is further $t$-dependence in the volume form
$\omega_{\mu\nu\rho\sigma\tau}$ and the inverse metric $g^{\tau\lambda}$. In the
basis where $t$ is used for the radial coordinate they are given by
\begin{align}
\label{eq:gIAdS}
g^{\lambda\tau} &= \operatorname{diag}(-e^{-2\phi(t)},e^{-2\phi(t)},e^{-2\phi(t)},e^{-2\phi(t)},t^2) \\
\label{eq:e2phi_t}
e^{2\phi(t)} &= \frac{1}{t^2}(1-t^2)(1-\lambda^2 t^6)^{1/3}e^{2 C_1} \\
\label{eq:AdS_volume_form}
\begin{split}
\omega_{\mu\nu\rho\sigma\tau} &= \frac{1}{t}e^{4\phi(t)}\epsilon_{\mu\nu\rho\sigma\tau}
= \frac{1}{t^5}\left(1-t^2\right)^2 \left(1-\lambda ^2 t^6\right)^{2/3} e^{4 C_1}\epsilon_{\mu\nu\rho\sigma\tau}.
\end{split}
\end{align}

%%%%%%%%%%%%%%%%%%%%%%%%%%%%%%%%%%%%%%%%%%%%%%%%%%%%
%
% S^5 Spherical Harmonics with SO(3)_diag Symmetry
%
%%%%%%%%%%%%%%%%%%%%%%%%%%%%%%%%%%%%%%%%%%%%%%%%%%%%
\section{\texorpdfstring{\boldmath$S^5$}{Five-Sphere} Spherical Harmonics with \texorpdfstring{\boldmath$\textup{SO}(3)_\text{diag}$}{diagonal SO(3)} Symmetry}%
\label{app:harmonics}
In this appendix we would like to list a subset of the $S^5$ scalar, vector, and
tensor spherical harmonics that are invariant under the $\textup{SO}(3)_\text{diag}
\subset \textup{SO}(3)\times \textup{SO}(3) \subset \textup{SO}(6)$ symmetry. These harmonics can be found
by solving the following defining Laplace eigenvalue equations under the
constraint that the solutions be $\textup{SO}(3)_\text{diag}$-invariant:
\begin{align}\label{eq:laplace_so3_diag}
0 &= \mathring\nabla^2 Y^{(k,m)} + k(k + 4)Y^{(k,m)} \\
0 &= \mathring\nabla^2 Y_{v\,n}^{(k,m)} + (k^2 + 6k + 4)Y_{v\,n}^{(k,m)} \\
0 &= \mathring\nabla^2 Y_{t\,[np]}^{(k,m)} + (k^2 + 6k + 3)Y_{t\,[np]}^{(k,m)} \\
0 &= \mathring\nabla^n Y_{v\,n}^{(k,m)} = \mathring\nabla^n Y_{t\,[np]}^{(k,m)} \\
k&=0,1,2,\dots \nonumber
\end{align}
The scalar harmonics are denoted by $Y^{(k,m)}$, the vector harmonics by
$Y_{v\,n}^{(k,m)}$, and the tensor harmonics by $Y_{t\,[np]}^{(k,m)}$. The
symbol $\mathring\nabla$ is the covariant derivative on the $S^5$, the indices
$n,p\in\{1,\dots,5\}$ in $Y_{v\,n}^{(k,m)}$ and $Y_{t\,[np]}^{(k,m)}$ refer to
the $S^5$ cotangent space, and the indices `$v$' and `$t$' stand for
``vector''and ``tensor''. The integer $m$ measures the degeneracy of the
harmonics for a given $k$. For scalar harmonics $m$ appears in the eigenvalue of
the following differential operator:
\begin{equation}
\frac{1}{\sin\phi}\frac{\partial}{\partial\phi}\left(\sin\phi\frac{\partial}{\partial\phi}Y^{(k,m)}\right)+m(m+1)Y^{(k,m)} = 0.
\end{equation}
In what follows we will normalize the harmonics so that
\begin{multline}
\int_{S^5}Y^{(k,m)} Y^{(k',m')} = \int_{S^5} Y_{v\,n}^{(k,m)}Y_v^{n(k',m')} = \int_{S^5} Y_{t\,[np]}^{(k,m)}Y_t^{[np](k',m')} \\
= \frac{\pi^3}{2^{k-1}(k + 1)(k + 2)}\delta^{k k'}\delta^{m m'}.
\end{multline}
The lowest scalar harmonics are given by
\begin{alignat}{2}
Y^{(0,0)} &= 1 \\
Y^{(2,0)} &= \frac{1}{\sqrt{6}} \cos 2\theta
&&= \frac{1}{\sqrt{6}}(u^2 - v^2) \\
Y^{(2,1)} &= \frac{1}{\sqrt{6}} \cos\phi \sin 2\theta
&&= \sqrt{\frac23} (u\cdot v) \\
Y^{(4,0)} &= \frac{1}{4\sqrt{15}} (2\cos 4\theta + 1)
&&= \frac{1}{4\sqrt{15}}(3u^4 - 10u^2 v^2 + 3v^4) \\
Y^{(4,1)} &= \frac{1}{2\sqrt{10}} \cos\phi \sin 4\theta
&&= \sqrt{\frac25} (u\cdot v)(u^2 - v^2) \\
Y^{(4,2)} &= \frac{1}{\sqrt{30}} \sin^2\theta \cos^2\theta \; (3\cos 2\phi + 1)
&&= \sqrt{\frac{2}{15}}[3 (u\cdot v)^2 - u^2 v^2] \\
Y^{(6,0)} &= \frac{1}{8\sqrt{7}} (\cos 2\theta + \cos 6\theta)
&&= \frac{1}{4\sqrt{7}}(u^2 - v^2)(u^4 - 6u^2 v^2 + v^4)\\
Y^{(6,1)} &= \frac{1}{8\sqrt{35}} \cos\phi \; (\sin 2\theta + 3\sin 6\theta)
&&= \frac{1}{2\sqrt{35}}(u\cdot v)(5 u^4 - 14 u^2 v^2 + 5v^4) \\
Y^{(6,2)} &= \frac{1}{2\sqrt{7}} \sin^2\theta \cos^2\theta \cos 2\theta \; (3\cos 2\phi + 1)
&&= \frac{1}{\sqrt{7}} (u^2 - v^2)[3(u\cdot v)^2 - u^2 v^2] \\
Y^{(6,3)} &= \frac{1}{\sqrt{35}} \sin^3\theta \cos^3\theta \cos\phi \; (5 \cos 2\phi - 1)
&&= \frac{2}{\sqrt{35}} (u\cdot v)[5(u\cdot v)^2 - 3 u^2 v^2].
\end{alignat}
The transformation from the angle coordinates $(\theta, \phi)$ to the embedding
coordinates $(\vec u, \vec v)$ is performed using $\cos^2\theta = u^2$,
$\sin^2\theta = v^2$, and $\sin\theta\cos\theta\cos\phi = u\cdot v$. The scalar
harmonics can also be converted to the $(w_1,w_2)$ basis using $w_1 = 2u^2 - 1 =
1 - 2v^2 = u^2 - v^2$ and $w_2 = 2u\cdot v$. One obtains the following
expressions
\begin{align}
Y^{(0,0)} &= 1 \\
Y^{(2,0)} &= \frac{w_1}{\sqrt{6}} \\
Y^{(2,1)} &= \frac{w_2}{\sqrt{6}} \\
Y^{(4,0)} &= \frac{4w_1^2 - 1}{4\sqrt{15}} \\
Y^{(4,1)} &= \frac{w_1 w_2}{\sqrt{10}} \\
Y^{(4,2)} &= \frac{w_1^2 + 3w_2^2 - 1}{2\sqrt{30}} \\
Y^{(6,0)} &= \frac{w_1(2w_1^2 - 1)}{4\sqrt{7}} \\
Y^{(6,1)} &= \frac{w_2(6w_1^2 - 1)}{4\sqrt{35}} \\
Y^{(6,2)} &= \frac{w_1(w_1^2 + 3w_2^2 - 1)}{4\sqrt{7}} \\
Y^{(6,3)} &= \frac{w_2(3w_1 + 5w_2^2 - 3)}{4\sqrt{35}}
\end{align}
The lowest vector harmonics that are used in the text are given by
\begin{align}
Y_v^{(0,0)} &= \cos\theta \sin\theta \sin\phi\,(2\sigma^1 + d\phi) - \cos\phi \; d\theta &&= u^i dv^i - v^i du^i \\
Y_v^{(1,0)} &= \frac{2}{\sqrt{3}}\sin\theta \cos^2\theta \sin\phi \; \sigma^2 &&= \frac{2}{\sqrt{3}}(\epsilon^{ijk}v^i u^j du^k) \\
Y_v^{(1,1)} &= \frac{2}{\sqrt{3}}\sin^2\theta \cos\theta \sin\phi\,(\cos\phi \; \sigma^2 - \sin\phi \; \sigma^3) &&= \frac{2}{\sqrt{3}}(\epsilon^{ijk}v^i u^j dv^k).
\end{align}
We can change the $(\theta,\phi)$ coordinates to $(w_1,w_2)$ and re-write the
harmonics as follows
\begin{align}
Y_v^{(0,0)} &= \sqrt{\zeta} \; \sigma^1 + \frac14\left(\frac{1+w_1}{2}\right)^{-1}w_2 \; dw_1 - \frac12 \; dw_2 \\
Y_v^{(1,0)} &= \sqrt{\frac13}\left(\frac{1+w_1}{2}\right)^{1/2}\sqrt{\zeta} \; \sigma^2 \\
Y_v^{(1,1)} &= \sqrt{\frac{1}{12}}\left(\frac{1+w_1}{2}\right)^{-1/2}(w_2\sqrt\zeta\sigma^2 - \zeta\sigma^3).
\end{align}
Note that we view the vector harmonics as one-forms $Y_v^{(k,m)}$ on the $S^5$
cotangent space. Their components $Y_{v\,n}^{(k,m)}$ can be read off by choosing
a basis of one-forms and expanding the vector harmonics in that basis.

The lowest tensor harmonics corresponding to $k=0$ are given by the following
two-forms on the $S^5$ cotangent space:
\begin{align}
Y_t^{(0,0)} &= \epsilon^{ijk} (u^i du^j \wedge du^k) \\
Y_t^{(0,1)} &= \frac{1}{\sqrt3}\epsilon^{ijk} (v^i du^j \wedge du^k + 2 u^i du^j \wedge dv^k) \\
Y_t^{(0,2)} &= \frac{1}{\sqrt3}\epsilon^{ijk} (u^i dv^j \wedge dv^k + 2 v^i dv^j \wedge du^k) \\
Y_t^{(0,3)} &= \epsilon^{ijk} (v^i dv^j \wedge dv^k).
\end{align}
We can change to the $(w_1, w_2)$ basis as before and write the tensor harmonics
in terms of the vector harmonics as follows:
\begin{align}
Y_t^{(0,0)} &= \frac{\sqrt3}{\zeta}(1+w_1) \left(Y_v^{(0,0)}-\frac{w_2 dw_1}{2 (1+w_1)}+\frac{dw_2}{2}\right)\wedge Y_v^{(1,0)}\\
Y_t^{(0,1)} &= \frac{1}{\zeta}(1+w_1) \left(Y_v^{(0,0)}-\frac{w_2 dw_1}{2 (1+w_1)}+\frac{dw_2}{2}\right)\wedge Y_v^{(1,1)} + \frac{1}{\zeta}(2 w_2 Y_v^{(0,0)}-dw_1)\wedge Y_v^{(1,0)} \\
Y_t^{(0,2)} &= \frac{1}{\zeta}(1-w_1)\left(Y_v^{(0,0)}-\frac{w_2 dw_1}{2(1-w_1)}-\frac{dw_2}{2}\right)\wedge Y_v^{(1,0)}+\frac{1}{\zeta}(2 w_2 Y_v^{(0,0)}-dw_1)\wedge Y_v^{(1,1)} \\
Y_t^{(0,3)} &= \frac{\sqrt{3}}{\zeta}(1-w_1) \left(Y_v^{(0,0)}-\frac{w_2 dw_1}{2 (1-w_1)}-\frac{dw_2}{2}\right)\wedge Y_v^{(1,1)}.
\end{align}

%%%%%%%%%%%%%%%%%%%%%%%%%%%%%%%%%%%%%%%%%%%%%%%%%%%%
%
% UV Asymptotics of the GPPZ uplift
%
%%%%%%%%%%%%%%%%%%%%%%%%%%%%%%%%%%%%%%%%%%%%%%%%%%%%
\section{UV Asymptotics of the GPPZ uplift}%
\label{app:asymptotics}

In this appendix we collect the UV expansion of the various fields in our
solution. Contrary to main text, we keep the explicit dependence on the angles
$\varphi$ and $\omega$ in~\eqref{angles}. When possible we express the results
on the basis of $S^5$ harmonics of Appendix~\ref{app:harmonics}.

\paragraph{Dilaton/axion}
The expansion of the axio-dilaton field up to terms of order $1/r^4$ is
\begin{eqnarray}
\label{app:axiodilUV}
B &=& \frac{1 + i \tau}{1- i \tau} \sim  - \frac{m_0^2}{r^2} \frac{ 4 \sqrt{2} }{3 \sqrt{3} }e^{-2 i \varphi} (Y^{(2,0)} + i Y^{(2,1)} ) \nonumber \\
&& + \frac{1}{r^4} \left[ \frac{ 2 \sqrt{2} }{3} m_0 \sigma_0 e^{- i ( \varphi - \omega)} ( Y^{(2,0)} - i Y^{(2,1)} ) + \frac{4 \sqrt{2}}{9 \sqrt{3}} m_0^4 e^{-2 i \varphi}  (Y^{(2,0)} + i Y^{(2,1)}) \right]
\end{eqnarray}
with $m_0$ and $\sigma_0$ given in~\eqref{eq:gppz_asymptotics}. The first terms
of the expansions of the dilaton and axion are
\begin{eqnarray}
e^\Phi &=& 1 + \frac{ m_0^2}{r^2} A_1  +  \frac{1 }{r^4} ( m_0 \sigma_0 A_2 + m_0^4  A_3 )  \\
C_0 &=& \frac{m_0^2}{r^2} B_1 + \frac{1 }{r^4} (m_0 \sigma_0 B_2 + m_0^4 B_3 )
\end{eqnarray}
where the coefficients $A_i$ and $B_i$ can be expressed in terms of scalar
harmonics as
\begin{eqnarray}
A_1 &=& - \frac{2 \sqrt{2}}{\sqrt{3}}  ( \sin 2\varphi  Y^{(2,1)}  +  \cos 2 \varphi Y^{(2,0)} ) \nonumber \\
B_1 &=& \frac{2 \sqrt{2}}{\sqrt{3}}  ( \cos 2\varphi Y^{(2,1)} -  \sin 2 \varphi Y^{(2,0)})  \nonumber \\
A_2 &=& \sqrt{2}  ( \cos( \varphi - \omega) Y^{(2,0)} -  \sin (\varphi - \omega) Y^{(2,1)} )  \nonumber \\
B_2 &=& \sqrt{2} ( \sin ( \varphi - \omega) Y^{(2,0)} +  \cos (\varphi - \omega) Y^{(2,1)} ) \nonumber \\
A_3 &=& \frac{1}{9}Y^{(0,0)} +  \frac{2 \sqrt{2}}{3 \sqrt{3}} ( \sin 2\varphi  Y^{(2,1)}  -  \cos 2 \varphi Y^{(2,0)} ) +
\frac{4 \sqrt{5}}{9 \sqrt{3}} (Y^{(4,0)} - \sqrt{2} Y^{(4,2)})   \nonumber \\
B_3 &=&  -  \frac{2 \sqrt{2}}{3 \sqrt{3}} ( \sin 2\varphi  Y^{(2,0)}  +  \cos 2 \varphi Y^{(2,1)} ) \nonumber \\
&& -
\frac{4 \sqrt{10}}{9 \sqrt{3} } (\sqrt{2} \sin 4 \varphi Y^{(4,0)}  - \sqrt{3} \cos 4 \varphi Y^{(4,1)}  - \sin 4 \varphi Y^{(4,2)}) \, .
\end{eqnarray}

\paragraph{Metric}
The large $r$ behaviour of the internal five-dimensional metric is, up to order
$1/r^4$
\begin{eqnarray}
d s^2_5 &=&  (d u_i)^2 + ( d v_i)^2 +  \frac{1}{r^2} ( A_{ij} d u_i d u_j + C_{ij}  d v_i d v_j + E_{ij} d u_i d v_j) \nonumber \\
&& + \frac{1}{r^4}  ( B_{ij}  d u_i d u_j + D_{ij}  d v_i d v_j + F_{ij}  du_i dv_j  )
\end{eqnarray}
with
\begin{eqnarray}
A_{ij} &=& - \frac{m_0^2}{6 } ( 3 + 4 (u^2 - v^2))  \delta_{ij} +   \frac{m_0^2}{3 }  v_i v_j \ \\
B_{ij}&=& \left[   \frac{m_0^4}{72} (7 + 40 (u^2 - v^2) + 96 (u \cdot v)^2 + 24 (u^2 - v^2)^2) \right. \nonumber \\
&& \left. + \frac{1}{\sqrt{3}} m_0 \sigma_0 ( \cos (\varphi + \omega) (u^2 -v^2) + 2 \sin (\varphi + \omega) (u \cdot v)) \right] \delta_{ij} \nonumber \\
&& + \frac{1}{\sqrt{3}} m_0 \sigma_0  [1 - \cos (\varphi + \omega) ] u_i u_j +  \left(  \frac{m_0^4}{6 } + \frac{1}{\sqrt{3}} \sigma_0 m_0 \cos (\varphi + \omega) \right)  v_i v_j \ \nonumber \\
&&+  \frac{1}{\sqrt{3}} m_0 \sigma_0 \sin (\varphi + \omega) (u_i v_j + v_i u_j) \\
C_{ij} &=&  - \frac{m_0^2}{6 } ( 3 - 4 (u^2 - v^2))  \delta_{ij}  +  \frac{m_0^2}{3 }  u_i u_j  \\
D_{ij} &=&  \left[  \frac{m_0^4}{72} (7 - 40 (u^2 - v^2) + 96 (u \cdot v)^2 + 24 (u^2 - v^2)^2) \right. \nonumber \\
&& \quad \left. - \frac{1}{\sqrt{3}} m_0 \sigma_0 ( \cos (\varphi + \omega) (u^2 -v^2) + 2 \sin (\varphi + \omega) (u \cdot v)) \right] \delta_{ij} \nonumber \\
&& +  \frac{1}{\sqrt{3}} m_0 \sigma_0  [1 - \cos (\varphi + \omega) ] v_i v_j +  \left(  \frac{m_0^4}{6}  + \frac{1}{\sqrt{3}} \sigma_0 m_0 \cos (\varphi + \omega) \right)  u_i u_j \ \nonumber \\
&& - \frac{1}{\sqrt{3}} m_0 \sigma_0 \sin (\varphi + \omega) (u_i v_j + v_i u_j)  \\
E_{ij}&=& \frac{m_0^2}{3} ( -8 (u \cdot v) \delta_{ij} - 8 u_i v_j + 6 v_i u_j) \\
F_{ij} &=& \left[ \frac{20 m_0^4}{9} ( u \cdot v) + \frac{2}{\sqrt{3}} m_0 \sigma_0 (\sin (\varphi + \omega) (u^2 - v^2) - 2 \cos (\varphi + \omega) (u \cdot v)) \right] \delta_{ij} \nonumber \\
&& + \frac{ 4 m_0^4}{9} \left[ 1+ \frac{3\sqrt{3} \sigma_0}{ 2 m_0^3} \right] u_i v_j +\frac{m_0^4}{9 } \left[ -7 + \frac{6 \sqrt{3} \sigma_0}{ m_0^3} \right] v_i u_j \nonumber \\
&& - \frac{2}{\sqrt{3}} m_0 \sigma_0 \sin (\varphi + \omega) (u_i u_j - v_i v_j)
\end{eqnarray}

\paragraph{Two-form potentials}
For generic values of the angles $\varphi$ and $\omega$, the two-form potentials
are
\begin{eqnarray}
\label{app:UVas2pot}
C_1 &=&  -  \frac{1}{2} \epsilon_{ijk}  \left[ \frac{ a_-(u,v) }{\sqrt{3} } \frac{m_0}{r}  -  \frac{ d_-(v,u) }{2} \frac{\sigma_0}{r^3} \right. \nonumber \\
&& \left. - \frac{2 c_+(v,u) +  2 ( 3 f^1_- u_i  + f^2_- v_i ) (u^2 - v^2) - 4 [  3 f^2_+ u_i - f^1_+ v_i ] ( u \cdot v) ) }{3 \sqrt{3}} \frac{m_0^3}{r^3} \right] d u_j \wedge d u_k  \nonumber \\
& & - \frac{1}{2} \epsilon_{ijk}  \left[ - \frac{b_-(v,u)}{\sqrt{3} } \frac{m_0}{r}  +  \frac{d_-(v,u)}{2} \frac{\sigma_0}{r^3} \right. \nonumber \\
 && \left.
 + \frac{2 c_-(v,u) - 2 ( f^1_- u_i + 3 f^2_- v_i ) (u^2 - v^2)  + 4 (f^2_+u_i - 3 f^1_+ v_i ) ( u \cdot v)}{3 \sqrt{3}} \frac{m_0^3}{r^3}  \right] d v_j \wedge d v_k  \nonumber \\
 &&+ \epsilon_{ijk}  \left[  \frac{c_-(u,v)}{\sqrt{3} } \frac{m_0}{r}  - \frac{d_-(u,v)}{2} \frac{\sigma_0}{r^3} \right. \\
 && \left. - \frac{2 c_-(u,v)  - 2 (f^2_- u_i + f^1_- v_i ) (u^2 - v^2) - 4 ( f^1_+ u_i - f^2_+ v_i ) ( u \cdot v) }{3 \sqrt{3}} \frac{m_0^3}{r^3} \right] d u_j \wedge d v_k \nonumber \\
C_2 &=&  -  \frac{1}{2} \epsilon_{ijk}  \left[ \frac{b_+(u,v)}{\sqrt{3} } \frac{m_0}{r}  -  \frac{d_+(u,v)}{2} \frac{\sigma_0}{r^3} \right. \nonumber \\
&& \left. - \frac{2 c_- (u,v) - 2 ( 3 g^1_+ u_i  + g^2_+ v_i ) (u^2 - v^2)  - 4 (  3 g^2_-u_i - g^1_- v_i ) ( u \cdot v) }{3 \sqrt{3}} \frac{m_0^3}{r^3}  \right] d u_j \wedge d u_k  \nonumber \\
& & - \frac{1}{2} \epsilon_{ijk}  \left[ \frac{a_+(v,u)}{\sqrt{3} } \frac{m_0}{r}  + \frac{d_+(u,v)}{2} \frac{\sigma_0}{r^3} \right. \nonumber \\
&& \left. + \frac{2 c_- (u,v) +2 (g^1_+ u_i + 3 g^2_+ v_i ) (u^2 - v^2) +4 (g^2_- u_i - 3 g^1_- v_i ) ( u \cdot v) }{3 \sqrt{3}} \frac{m_0^3}{r^3}
 \right] d v_j \wedge d v_k  \nonumber \\
 &&- \epsilon_{ijk}  \left[ \frac{c_+(v,u)}{\sqrt{3} } \frac{m_0}{r} +  \frac{d_-(v,u)}{2} \frac{\sigma_0}{r^3} \right.  \\
&& \left. - \frac{2 c_+(v,u)   - 2 ( g^2_+ u_i + g^1_+ v_i ] (u^2 - v^2) + 4 (g^1_- u_i - g^2_- v_i ) ( u \cdot v) }{3 \sqrt{3}} \frac{m_0^3}{r^3} \nonumber
 \right] d u_j \wedge d v_k
\end{eqnarray}
where
\begin{align}
& a^\pm (x,y) = 3 \sin \varphi x_i \pm \cos \varphi y_i  & \qquad  c^\pm (x,y) = \cos \varphi x_i \pm \sin \varphi y_i  \nonumber \\
& b^\pm (x,y) = 3 \cos \varphi x_i \pm \sin \varphi y_i  & \qquad  d^\pm(x,y) = \cos \omega x_i \pm \sin \omega y_i \nonumber
\end{align}
and
\begin{eqnarray}
&& f^1_\pm =\sin \frac{\varphi + \omega}{4} \cos \frac{3\varphi - \omega}{4} \pm \cos 3 \frac{\varphi + \omega}{4} \sin \frac{3\varphi - \omega}{4} \nonumber \\
&& f^2_\pm = \cos \frac{\varphi + \omega}{4} \cos \frac{3\varphi - \omega}{4} \pm \sin 3 \frac{\varphi + \omega}{4} \sin \frac{3\varphi - \omega}{4} \nonumber \\
&& g^1_\pm =\cos 3 \frac{\varphi + \omega}{4} \cos \frac{3\varphi - \omega}{4} \pm \sin \frac{\varphi + \omega}{4} \sin \frac{3\varphi - \omega}{4} \nonumber \\
&& g^2_\pm = \sin 3 \frac{\varphi + \omega}{4} \cos \frac{3\varphi - \omega}{4} \pm \cos \frac{\varphi + \omega}{4} \sin \frac{3\varphi - \omega}{4}
\end{eqnarray}

%%%%%%%%%%%%%%%%%%%%%%%%%%%%%%%%%%%%%%%%%%%%%%%%%%%%
%
% One Field Truncations
%
%%%%%%%%%%%%%%%%%%%%%%%%%%%%%%%%%%%%%%%%%%%%%%%%%%%%
\section{One Field Truncations}%
\label{app:1-field}

\subsection{The Case \texorpdfstring{$m(y)=0$}{m(y)=0}}

For the truncation $m(y)\equiv 0$ the uplift of the resulting $D=5$ theory has
been given by Gubser, Herzog, Pufu and Tesileanu~\cite{Gubser:2009qm}, see also
~\cite{Cassani:2010uw, Gauntlett:2010vu}. In the language of our uplift this
corresponds to setting $\nu \equiv 1$. After applying this to our uplift, and
using the usual definition $\mu = e^\sigma$ the warp factor $\Delta$ is simply
given by
\begin{eqnarray}
\Delta &\to (\cosh\sigma)^{-3/2}
\;,
\end{eqnarray}
and the IIB metric takes the form
\begin{eqnarray}
ds_{10}^2 &=& \cosh\sigma\left[(\cosh\sigma)^{-2/3} e^{2y}\eta_{\mu\nu}dx^\mu dx^\nu + dy^2\right]
\\
&&{}
+ \frac{1}{\cosh\sigma}\left[du^i du^i + dv^i dv^i + \left(\sinh\sigma Y_v^{(0,0)}\right)^2\right] \nonumber\\
&=& \cosh\sigma\left[\left(\frac{\sigma_0}{2\sinh\sigma}\right)^{2/3}\eta_{\mu\nu}dx^\mu dx^\nu + \left(\frac{2d\sigma}{3\sinh(2\sigma)}\right)^2 + \frac{d\Omega_5^2}{\cosh^2\sigma} + \left(\tanh\sigma Y_v^{(0,0)}\right)^2\right]
\;.
\nonumber
\end{eqnarray}
In the second line in the metric we change the radial variable from $y$ to
$\sigma$. The field $\sigma(y)$ is monotonic in $y$ for $y\in\{C_2,\infty\}$,
and is therefore invertible. The range of $\sigma$ is $\sigma\in\{0,\infty\}$,
where $\sigma = 0$ corresponds to the AdS boundary, and $\sigma = \infty$ to the
singularity. The Ricci scalar takes the simple form
\begin{eqnarray}
R &=& 18\sinh\sigma\tanh\sigma
\;.
\end{eqnarray}
The asymptotic behaviour of the metric for $\sigma \to \infty$, which is the
expansion around the singularity is the following
\begin{align}
ds_{10}^2 &= \left(\frac12 e^\sigma\right)^{1/3}\eta_{\mu\nu}dx^\mu dx^\nu + \frac19\left(\frac12 e^\sigma\right)^{-3}d\sigma^2 + \frac12 e^\sigma d\Omega_5^2 + \frac12 e^\sigma\left(Y_v^{(0,0)}\right)^2 + \dots \\
&=\rho^{2/3}\eta_{\mu\nu}dx^\mu dx^\nu + \left(\frac23 \frac{d\rho}{\rho^4}\right)^2 + \rho^2\left[d\Omega_5^2 + \left(Y_v^{(0,0)}\right)^2\right] + \dots
\;,
\end{align}
where we defined $\rho^2 = \frac12 e^\sigma$. The asymptotic behaviour of the
metric for $\sigma \to 0$, which is the expansion around the conformal boundary
of the AdS, is the following
\begin{align}
ds_{10}^2 &\to \left(\frac{\sigma_0}{2\sigma}\right)^{2/3}\eta_{\mu\nu}dx^\mu dx^\nu + \left(\frac{d\sigma}{3\sigma}\right)^2 + d\Omega_5^2 \\
&\qquad{} + \sigma^2\left[\frac{7}{18}\left(\frac{\sigma_0}{2\sigma }\right)^{2/3}\eta_{\mu\nu}dx^\mu dx^\nu -\frac{5}{54}d\sigma^2 - \frac12 d\Omega_5^2 + \left(Y_v^{(0,0)}\right)^2\right] + \dots \\
&= \frac{\eta_{\mu\nu}dx^\mu dx^\nu + dz^2}{z^2} + d\Omega_5^2 \\
&\qquad{} + \frac{\sigma_0^2}{4}z^6\left[\frac{7}{18}\frac{1}{z^2}\eta_{\mu\nu}dx^\mu dx^\nu - \frac{5}{24}\sigma_0^2 z^4 dz^2 - \frac12 d\Omega_5^2 + \left(Y_v^{(0,0)}\right)^2\right] + \dots
\end{align}
where we have substituted $z = \left(\frac{\sigma_0}{2\sigma}\right)^{-1/3}$.

In this truncation, the dilaton/axion matrix reduces to the identity, and the
two form is given by
\begin{eqnarray}
C_{mn\,\alpha} &=&{\cal S}_{\alpha}{}^a\,{\rm C}_{mn\,a}
\;,
\end{eqnarray}
with
\begin{eqnarray}
{\rm C}_1 &=&
-\frac14\,(1-\tanh\sigma) \,\varepsilon^{ijk} \,
\Big(
V^i\left( \Theta^j \wedge\Theta^k
 -
 \Lambda^j \wedge\Lambda^k\right)
%\nonumber\\
%&&{}\qquad\qquad
+ 2\, U^i \Theta^j \wedge \Lambda^k
 \Big)
 \;,
\nonumber\\[1ex]
{\rm C}_2 &=&-{\rm C}_1\Big|_{U^i\leftrightarrow V^i, \Theta^i\leftrightarrow \Lambda^i}
\;.
\end{eqnarray}

\subsection{The Case \texorpdfstring{$\sigma(y)=0$}{sigma(y)=0}}

Another consistent truncation to one scalar corresponds to setting
$\sigma(y)=0$, i.e.\ $\mu \equiv 1$. In this case, the IIB metric takes the form
\begin{eqnarray}
ds_{\rm IIB}^2 &=&
 \Delta^{-2/3}\left({g}_{\mu\nu}(x) \, dx^\mu dx^\nu
 + \Delta^{8/3}\, d\hat{s}_5^2 \right)
 \;,
 \label{metricIIBexplicits0}
\end{eqnarray}
with the warp factor $\Delta$ and the internal metric $d\hat{s}_5^2$ given by
\begin{eqnarray}
\Delta^{-8/3} &=&
\frac{ \left(1+ \nu ^2\right)^3 \left(1+\nu ^6\right)}{16\,  \nu ^6}
  + \frac{U^2\,V^2-(U\cdot V)^2}{16\, \nu ^8}\,
  \left(1-\nu ^4\right)^4
  \;,
  \label{warps0}
\\[2ex]
d\hat{s}_5^2 &=&
\frac{ \left(1+\nu ^4\right) \left(1+\nu ^2\right)^2 }{8\, \nu ^4}
 \left( \Theta^i\Theta^i + \Lambda^i\Lambda^i \right)
\nonumber\\
&&{}
  -\frac{ \left(1-\nu ^4\right)^2}{8\, \nu ^4}
  \left((U^2-V^2)\,(\Theta^i\Theta^i - \Lambda^i\Lambda^i)+4\,(U\cdot V) \,\Theta^i\Lambda^i \right)
\nonumber\\
&&{}
+\frac{\left(1-\nu^4\right) (1- \nu^2)
\left(1+\nu ^6\right)}{16\, \nu ^6}
 \left((V^i\Theta^i)(V^j\Theta^j)  + (U^i\Lambda^i )(U^j \Lambda^j)\right)
\nonumber\\
&&{}
   + \frac{ \left(1 -\nu ^4\right) \left(1+ \nu ^2\right)
   \left(1-\nu ^6\right)}{8\, \nu ^6}\, (V^i \Theta^i)(U^j \Lambda^j)
\nonumber\\
&&{}
  -\frac{(1-\nu^4)(1-\nu^8)}{4\,\nu^6} \, (U^i \Theta^i) (V^j \Lambda^j)
  \;,
  \label{G5s0}
\end{eqnarray}
respectively.
The dilaton/axion matrix $m_{\alpha\beta}$ is given by
\begin{eqnarray}
m_{\alpha\beta} &=&
\Delta^{4/3}\,{\cal S}_\alpha{}^a {\cal S}_\beta{}^b\,{\rm m}_{ab}
\;,
\label{dilaxs0}
\end{eqnarray}
with the ${\rm SO}(2)$ rotation matrix ${\cal S} = e^{\frac{3i}{4}\,\varphi\,\sigma_2}$
and the matrix ${\rm m}_{ab}$ given by
\begin{eqnarray}
{\rm m}_{11}&=&
\frac{1}{8\, \nu ^4}
\Big(
\left(1+\nu ^4\right) \left(1+\nu ^2\right)^2
+\left(1-\nu ^4\right)^2 \left(U^2-V^2\right)
\Big)
  \;,\nonumber\\
 {\rm m}_{12} &=&
\frac{ \left(1-\nu ^4\right)^2 }{4\, \nu ^4}
  \left(U\cdot V\right)
\;,
 \nonumber\\
{\rm m}_{22} &=&
\frac{1}{8\, \nu ^4}
\Big(
\left(1+\nu ^4\right) \left(1+\nu ^2\right)^2
-\left(1-\nu ^4\right)^2 \left(U^2-V^2\right)
\Big)
\;.
\label{mms0}
\end{eqnarray}
The two-form doublet is given by
\begin{eqnarray}
C_{mn\,\alpha} &=& \Delta^{8/3}\,{\cal S}_{\alpha}{}^a\,{\rm C}_{mn\,a}
\;,
\end{eqnarray}
with
\begin{eqnarray}
{\rm C}_1 &=&
-\frac{1}{8\, \nu ^4}\,{\rm m}_{11}\,\varepsilon^{ijk} \,
\Big(
\left(1- \nu ^2\right)\left(1+\nu ^6\right)
V^i\, \Theta^j \wedge\Theta^k
 +\left(1+ \nu^2\right) \left(1-\nu ^6\right)
 V^i\, \Lambda^j \wedge\Lambda^k
\nonumber\\
&&{}\qquad\qquad
+ 2\,\nu^2\left(1-\nu ^4\right) U^i \, \Theta^j \wedge \Lambda^k
 \Big)
 \nonumber\\
 &&{}
 + \frac{1}{8\, \nu ^4}\,{\rm m}_{12}\,\varepsilon^{ijk} \,
\Big(
\left(1+ \nu ^2\right)\left(1-\nu ^6\right)
U^i\, \Theta^j \wedge\Theta^k
 +\left(1- \nu^2\right) \left(1+\nu ^6\right)
 U^i\, \Lambda^j \wedge\Lambda^k
\nonumber\\
&&{}\qquad\qquad
+ 2\,\nu^2 \left(1- \nu ^4\right) V^i \, \Theta^j \wedge \Lambda^k
 \Big)
 \;,
\nonumber\\[1ex]
{\rm C}_2 &=&-{\rm C}_1\Big|_{U^i\leftrightarrow V^i, \Theta^i\leftrightarrow \Lambda^i}
\;.
\label{C2s0}
\end{eqnarray}
Finally, the non-vanishing components of the 5-form field strength are given by
\begin{eqnarray}
F_{\mu\nu\rho\sigma\tau} &=&
-\frac1{3}\,\omega_{\mu\nu\rho\sigma\tau}\,V_{\rm pot}
\;,
\nonumber\\
F_{\mu\nu\rho\sigma m} &=&
\frac{1}{8} \,\omega_{\mu\nu\rho\sigma\lambda}\,
{\cal J}^\lambda
\left(
U^i\Lambda^i-V^i\Theta^i
\right)_m
\;,\nonumber\\
F_{\mu\nu\rho mn}
&=&
-\frac{1}{12}\,\omega_{\mu\nu\rho\sigma\tau}\, F^{\sigma\tau}\,
 (\Theta^i\wedge \Lambda^i)_{mn}
\;,
\label{F5s0}
\end{eqnarray}
together with those related by self-duality of the IIB field strength.

%%%%%%%%%%%%%%%%%%%%%%%%%%%%%%%%%%%%%%%%%%%%%%%%%%%%
%
% BIBLIOGRAPHY
%
%%%%%%%%%%%%%%%%%%%%%%%%%%%%%%%%%%%%%%%%%%%%%%%%%%%%
\bibliographystyle{JHEP}
\providecommand{\href}[2]{#2}

\end{document}